\documentclass[prd,twocolumn,showpacs,preprintnumbers,amsmath,nofootinbib,amssymb,superscriptaddress,letter]{revtex4-1}

\usepackage{bbm,lmodern,amsmath,graphicx,epsfig,amssymb,amsfonts,dsfont,mathtools}
\usepackage{amsmath,amssymb,mathrsfs}
\usepackage{rotating}
\usepackage{bigstrut}
\usepackage{morefloats}
\usepackage{multirow}
\usepackage{longtable}
\usepackage{dcolumn}
\usepackage{placeins}
\usepackage{ulem}
\usepackage[dvipsnames]{xcolor}
\usepackage[utf8]{inputenc}
\usepackage{nicefrac} 

\usepackage{hyperref}
\hypersetup{
	colorlinks	=true,
	urlcolor	=blue,
	linkcolor	=blue,
	citecolor	=blue,
	pdftitle	={},
	pdfauthor	={D.~R\"onchen, M.~D\"oring, H.~Haberzettl, J.~Haidenbauer, U.-G.~Mei\ss ner, K. Nakayama},
	pdfsubject	={The impact of $\mathbf{K^+\Lambda}$ photoproduction on the resonance spectrum}
	}

\allowdisplaybreaks[1]

\newcommand{\be}{\begin{equation}}
\newcommand{\ee}{\end{equation}}
\newcommand{\ba}{\begin{eqnarray}}
\newcommand{\ea}{\end{eqnarray}}
\newcommand{\non}{\nonumber \\}
\newcommand{\po}{{\rm P}}
\newcommand{\npo}{{\rm NP}}

\begin{document}

\preprint{JLAB-THY-18-2640}

\title{The impact of $\mathbf{K^+\Lambda}$ photoproduction on the resonance spectrum}

\author{D.~R\"onchen}
\email{roenchen@hiskp.uni-bonn.de}
\affiliation{Helmholtz-Institut f\"ur Strahlen- und Kernphysik (Theorie) and Bethe Center for Theoretical
Physics,  Universit\"at Bonn, 53115 Bonn, Germany}

\author{M.~D\"oring}
\email{doring@gwu.edu}
\affiliation{Institute for Nuclear Studies and Department of Physics, The George Washington University,
Washington, DC 20052, USA}
\affiliation{Thomas Jefferson National Accelerator Facility, 12000 Jefferson Avenue, Newport News,VA, USA}

\author{U.-G.~Mei\ss ner}
\email{meissner@hiskp.uni-bonn.de}
\affiliation{Helmholtz-Institut f\"ur Strahlen- und Kernphysik (Theorie) and Bethe Center for Theoretical
Physics,  Universit\"at Bonn, 53115 Bonn, Germany}
\affiliation{Institute for Advanced Simulation, Institut f\"ur Kernphysik, J\"ulich Center for Hadron Physics and JARA HPC, Forschungszentrum J\"ulich, 
52425 J\"ulich, Germany}

\begin{abstract}
The J\"ulich-Bonn coupled-channel framework is extended to $K^+\Lambda$ photoproduction. The spectrum of nucleon and $\Delta$ resonances is extracted from simultaneous fits to several pion-induced reactions in addition to pion, eta and $K^+\Lambda$ photoproduction off the proton. 
More than 40,000 data points up to a center-of-mass energy of $E\sim 2.3$~GeV including recently measured double-polarization observables are analyzed. 
The influence of the $\gamma p\to K^+\Lambda$ channel on the extracted resonance parameters and the appearance of states not seen in other channels is investigated. The J\"ulich-Bonn model includes effective three-body channels and guarantees unitarity and analyticity, which is a prerequisite for a reliable determination of the resonance spectrum in terms of poles and residues.

\end{abstract}


\maketitle


\section{Introduction}

The excitation spectrum of baryons provides a connection between Quantum Chromodynamics and experiment in the low and medium energy regime of the strong interactions, where a perturbative treatment of QCD is not feasible. 
For many years, elastic and charge-exchange $\pi N$ scattering was the main source of information for studying the $N^*$ and $\Delta^*$ spectrum in the traditional partial-wave analyses~\cite{Hoehler1,Cutkosky:1979fy,Arndt:2006bf}. Compared to quark model predictions~\cite{Capstick:1993kb,Ronniger:2011td} or lattice simulations~\cite{Edwards:2011jj,     Edwards:2012fx, Lang:2012db, Engel:2013ig, Lang:2016hnn, Kiratidis:2016hda, Andersen:2017una}, however, the number of resonances seen in $\pi N$ scattering is much smaller, a situation referred to as the ``missing resonance problem"~\cite{Koniuk:1979vw}.  
In recent years, the experimental study of reactions other than $\pi N$ elastic scattering was given much attention at photon-beam facilities like ELSA, JLab or MAMI. High-quality data for cross sections, single- and double polarization observables are nowadays available for different final states~\cite{Crede:2013sze, Aznauryan:2012ba, Klempt:2009pi} and will in the near future allow the determination of the photoproduction amplitude from a ``complete experiment"~\cite{Barker:1975bp}, a set of eight carefully chosen observables that resolve all discrete ambiguities in pseudoscalar meson photoproduction up to an overall phase~\cite{Chiang:1996em,Keaton:1996pe}. 
Although experimental data with realistic uncertainties require more than eight observables~\cite{Sandorfi:2010uv,Ireland:2010bi,Nys:2015kqa}, it is possible to perform a truncated partial-wave analysis with less than eight~\cite{Wunderlich:2014xya,Workman:2016irf}.
In this respect, the photoproduction of $KY$ final states offers the advantage that the recoil polarization is accessible through the self-analyzing weak decay of the hyperon. A complete set which always includes beam-recoil or target-recoil measurements is, thus, easier to realize. 
See also Ref.~\cite{Nys:2016uel} in this context in which the question is addressed of how precise data have to be to discriminate between models.

Kaon photoproduction holds the promise to reveal resonances not seen in pion or eta production. The strangeness channels $K\Lambda$ and $K\Sigma$ might be the dominant decay modes of states that couple only weakly to $\pi N$ or $\eta N$, especially at energies farther away from the $\pi N$ threshold. 
The data situation for the reaction $\pi^-p\to K^0\Lambda$ is much better than in other pion-induced channels. In the coupled-channel fit of pion-induced reactions of Ref.~\cite{Ronchen:2012eg} the inclusion of the $K^0\Lambda$ final state data led to strong evidence of the $N(1710)1/2^+$ resonance. Yet, despite a small amount of data points available for the spin-rotation parameter $\beta$~\cite{Bell:1983dm}, the quality of the polarization data does not permit an unambiguous determination of the amplitude~\cite{Ronchen:2012eg}. 
The study of kaon photoproduction is, thus, a vital step towards establishing the baryon excitation spectrum and could contribute to solving the missing resonance puzzle. 

Theoretical studies of kaon photoproduction have been pursued using a variety of different approaches. The energy region not far away from threshold can be studied  in the framework of chiral perturbation theory~\cite{Steininger:1996xw}. Yet, as shown in Ref.~\cite{Bijnens:1985kj,Mai:2009ce} SU(3) ChPT converges rather slowly in the hadronic sector. Another method uses unitarized chiral interactions~\cite{Kaiser:1996js,Borasoy:2007ku}. In Ref.~\cite{Golli:2016dlj} a chiral quark model is applied to predict amplitudes for eta and kaon production. 
Single-channel isobar models~\cite{Mart:1999ed,Clymton:2017nvp,Skoupil:2018vdh} and multi-channel $K$-matrix approaches~\cite{Anisovich:2007bq,Anisovich:2014yza,Anisovich:2017pmi, Anisovich:2017ygb, Shklyar:2005xg,Cao:2013psa,Hunt:2018mrt} 
cover a broad energy range and are able to analyse a large amount of data. 
To this purpose, the real dispersive parts of the intermediate states are often neglected which allows for a flexible and effective parametrization of the amplitude although certain $S$-matrix principles
like three-body unitarity are impossible to implement.

Based on effective Lagrangians, dynamical coupled-channel (DCC) approaches preserve, or at least approximate, theoretical constraints of the $S$-matrix like two- and three-body unitarity, analyticity, left-hand cuts and complex branch points. 
This ensures a reliable determination of the resonance spectrum in terms of complex pole positions and residues. 
DCC approaches provide a particularly suited tool for a simultaneous analysis of multiple channels over a wide energy range.  A DCC analysis including kaon photoproduction was performed in Refs.~\cite{Kamano:2013iva,Kamano:2016bgm}, see also Refs.~\cite{JuliaDiaz:2006is, Suzuki:2009nj}. Other notable approaches to kaon photoproduction comprise kaon-MAID~\cite{Lee:1999kd}, the Regge-plus-resonance parametrization of Refs.~\cite{DeCruz:2011xi,DeCruz:2012bv} and the analysis of Ref.~\cite{Maxwell:2012zz} using an effective Lagrangian model.

In the present study we extend the J\"ulich-Bonn DCC approach to the $\gamma p\to K^+\Lambda$ channel. The J\"ulich-Bonn model was developed over the years~\cite{Doring:2009bi,Doring:2009yv,Doring:2010ap,
Ronchen:2012eg,Ronchen:2014cna} starting with Ref.~\cite{Schutz:1994ue} and includes in its most recent form the pion-induced reactions $\pi N\to \pi N$, $\pi^-p\to\eta n$, $K^0\Lambda$, $K^0\Sigma^0$, $K^+\Sigma^-$ and $\pi^+p\to K^+\Sigma^+$, in addition to pion and eta photoproduction off the proton~\cite{Ronchen:2015vfa}.
Recently, the J\"ulich-Bonn approach was also extended to the hidden-charm and hidden-beauty sector to explore the possibility of dynamically generated resonances in the 4~GeV and 11~GeV energy regime~\cite{Shen:2017ayv}.  

The paper is organized as follows: In Sec.~\ref{sec:formalism} we give a short introduction to the J\"ulich-Bonn framework. A detailed description of the hadronic interaction can be found in Ref.~\cite{Ronchen:2012eg} while the parametrization of the photoproduction amplitude is developed in Ref.~\citep{Ronchen:2014cna}. Sec.~\ref{sec:results} includes numerical details and fit results and the extracted resonance spectrum is discussed in Sec.~\ref{sec:spectrum}.

\section{Formalism}
\label{sec:formalism}

The J\"ulich-Bonn (J\"uBo) model was originally developed to describe $\pi N$ interaction. A simultaneous analysis of the reactions $\pi N\to \pi N$, $\eta N$, $K\Lambda$ and $K\Sigma$ was achieved in Ref.~\cite{Ronchen:2012eg}. The hadronic scattering potential is derived from an effective Lagrangian using time-ordered perturbation theory (TOPT) and is iterated in a Lippmann-Schwinger equation, which automatically ensures two-body unitarity. Two-to-three and three-to-three body unitarity is approximately fulfilled and the $\pi\pi N$ channels are parameterized as $\rho N$, $\sigma N$ and $\pi\Delta$. Those channels are included dynamically in the sense that the $\pi\pi$ and $\pi N$ subsystems fit the respective phase shifts~\cite{Doring:2009yv}.
The amplitude is inspired by the Amado model~\cite{Aaron:1969my} although the correct proof of three-body unitarity has been provided only recently~\cite{Mai:2017vot}. Moreover, there it was shown that the amplitude, although formulated in terms of isobars and spectators, can be entirely re-formulated in terms of on-shell two-body amplitudes, their continuation below two-body thresholds, and real-valued three-body forces. 

Note also that the large $\pi\pi N$ inelasticities in the light baryon sector represent one of the main obstacles to interpret lattice QCD calculations performed in a small cubic volume (see, e.g., Ref.~\cite{Lang:2016hnn}). As three-body unitarity identifies the imaginary parts of the amplitude in the infinite volume (when all particles are on-shell), it can be used to determine and correct for the leading power-law finite-volume effects arising from the three-body on-shell condition~\cite{Mai:2017bge}.

The J\"uBo approach respects analyticity; left-hand cuts and the correct structure of real and complex branch points~\cite{Ceci:2011ae} as well as the real, dispersive contributions of intermediate states are taken into account. More details on the analytic properties of the scattering amplitude are given in Ref.~\citep{Doring:2009yv}.

In Ref.~\cite{Huang:2011as} a field-theoretical description of pseudoscalar meson photoproduction that fulfils the generalized off-shell Ward-Takahashi identity and uses an earlier version of the J\"uBo model as final-state interaction is presented. 
In the present study we follow a different approach and approximate the photoproduction kernel by energy-dependent polynomials while the hadronic final-state interaction is given by the J\"uBo model in its current version~\cite{Ronchen:2012eg}. This semi-phenomenological framework is  more flexible than the technically rather involved covariant treatment of Ref.~\cite{Huang:2011as} and is especially suited to analyse the large amounts of data nowadays available for meson photoproduction. 
While no information on the underlying microscopic photoexcitation process can be gained, the good analytic properties of the hadronic $T$-matrix allow for a well defined extraction of the resonance spectrum. This approach is similar to the GWU/DAC CM12 framework of Ref.~\cite{Workman:2012jf}. Its integration into the J\"uBo formalism and the application in an analysis of pion photoproduction can be found in Ref.~\cite{Ronchen:2014cna}. In Ref.~\cite{Ronchen:2015vfa} the analysis was extended to eta photoproduction.

In the following we briefly describe the main ingredients of the framework. For a detailed description of the J\"uBo approach the reader is referred to Refs.~\cite{Ronchen:2012eg,Ronchen:2014cna}.

The hadronic scattering process of a meson and a baryon is described by the following scattering equation: 
\begin{multline}
T_{\mu\nu}(q,p',E_{\rm cm})=V_{\mu\nu}(q,p',E_{\rm cm}) \\
+\sum_{\kappa}\int\limits_0^\infty dp\,
 p^2\,V_{\mu\kappa}(q,p,E_{\rm cm})G^{}_\kappa(p,E_{\rm cm})\,T_{\kappa\nu}(p,p',E_{\rm cm}) \ .
\label{eq:scat}
\end{multline}
This equation is formulated in partial-wave basis and the indices $\mu$, $\nu$ and $\kappa$ denote the outgoing, incoming and intermediate meson-baryon channels, respectively. $E_{\rm cm}$ stands for the scattering energy in the center-of-mass frame while $q\equiv|\vec q\,|$ ($p^\prime\equiv|\vec p^{\,\prime}|$) indicates the modulus of the outgoing (incoming) three-momentum. Note that the latter can be on- or off-shell. 

In case of channels with stable particles the propagator $G_\kappa$ is of the form
\begin{equation}
G_\kappa(p,E_{\rm cm})=\frac{1}{E_{\rm cm}-E_a(p)-E_b(p)+i\epsilon}\;,
\label{eq:prop}
\end{equation}
where $E_a=\sqrt{m_a^2+p^2}$ and  $E_b=\sqrt{m_b^2+p^2}$ are the on-mass-shell energies of the
intermediate particles $a$ and $b$ in channel $\kappa$ with masses $m_a$ and $m_b$. Eq.~(\ref{eq:prop}) applies to $\kappa=\pi N$, $\eta N$, $K\Lambda$, or $K\Sigma$; for channels with unstable particles, i.e. $\rho N$, $\sigma N$ and $\pi\Delta$, the propagator has a more complex form~\cite{Doring:2009yv,Krehl:1999km}.

The scattering potential $V_{\mu\nu}$ is constructed of $t$- and $u$-channel exchanges of known mesons and baryons and $s$-channel pole terms that account for genuine resonances (see Ref.~\cite{Ronchen:2012eg} for a complete list of $t$- and $u$-channel exchanges). In addition, contact diagrams are included that absorb physics beyond the explicit processes. Those contact interaction preserve the analytic properties ensured by the $t$-, $u$- and $s$-channel interactions. The potential can, thus, be decomposed into three parts,
\begin{eqnarray}
V_{\mu\nu}&=&V^\npo_{\mu\nu}+V^\po_{\mu\nu}+ V_{\mu\nu}^{{\rm CT}}  \nonumber \\&\equiv & V^\npo_{\mu\nu}+\sum_{i=0}^{n} 
\frac{\gamma^a_{\mu;i}\,\gamma^c_{\nu;i}}{E_{\rm cm}-m_i^b}+\frac{1}{m_N}\gamma^{{\rm CT};a}_{\mu}\gamma^{{\rm CT};c}_\nu \;. 
\label{eq:vstuct}
\end{eqnarray}
The non-pole part of the potential, $V^\npo$, comprises all $t$- and $u$-channel exchange diagrams, while $V^\po$ includes all $s$-channel resonances and the vertex functions $\gamma^c_{\nu;i}$ ($\gamma^a_{\mu;i}$) describe the creation (annihilation) of a resonance $i$ in channel $\nu$ ($\mu$) with bare mass $m_i^b$. A compilation of all exchange processes included in the approach as well as explicit formulas for exchange potentials and resonance vertex functions are given in Refs.~\cite{Ronchen:2012eg,Doring:2010ap}. The vertex functions of the contact diagrams $\gamma^{{\rm CT};a}_{\mu}$ ($\gamma^{{\rm CT};c}_\nu$) have the same functional form as the resonance vertex functions. All parts of the scattering potential include free parameters that are fitted to data. Details will be given in Sec.~\ref{sec:numerics}.

Similar to the potential $V$, the scattering matrix of Eq.~(\ref{eq:scat}) can be decomposed into a pole and a non-pole part, a decomposition widely used in the literature, 
 \begin{equation}
T_{\mu\nu}=T_{\mu\nu}^\po+T_{\mu\nu}^\npo 
\label{eq:deco1} 
\end{equation} 
with
 \begin{equation}
T^\npo_{\mu\nu}=V_{\mu\nu}^\npo+\sum_\kappa V_{\mu\kappa}^\npo G^{}_{\kappa} T_{\kappa\nu}^\npo \ .
\label{eq:tnp} 
\end{equation} 

The pole part $T^{\rm P}$ can be evaluated from the non-pole $T^{\rm NP}$:
\be
T^{\po}_{\mu\nu}=\Gamma^a_{\mu;i}\, D^{}_{ij} \, \Gamma^c_{\nu;j}
\label{tpo}
\ee
with the resonance propagator $D_{ij}$ and the dressed creation (annihilation) vertex $\Gamma_{\mu;i}^c$ ($\Gamma_{\mu;i}^a$), 
\begin{eqnarray}
\Gamma_{\mu;i}^c	&=&\gamma^c_{\mu;i}+\sum_\nu  \gamma^c_{\nu;i}\,G^{}_\nu\,T_{\nu\mu}^\npo \ , \non
\Gamma_{\mu;i}^a	&=&\gamma^a_{\mu;i}+\sum_\nu  T_{\mu\nu}^\npo\,G^{}_\nu\,\gamma^a_{\nu;i} \ .
\label{eq:dressed}
\end{eqnarray}
The indices $i$ and $j$ label the $s$-channel states or a contact diagram in a given partial wave. In case of  two $s$-channel resonances with bare masses $m^b_1$ and $m^b_2$ (indices $i,j\in \{1,2\}$) plus one contact term (indices $i,j=3$), the dressed vertex functions and the resonance propagator are of the form
\begin{eqnarray}
\Gamma^a_\mu &=&(\Gamma^a_{\mu;1},\Gamma^a_{\mu;2},\Gamma^a_{\mu;3}), \quad
\Gamma^c_\mu=\left(
\begin{matrix}
\Gamma^c_{\mu;1}\\
\Gamma^c_{\mu;2}\\
\Gamma^c_{\mu;3}
\end{matrix}
\right), \non
D^{-1}&=&\left(\begin{matrix}
E_{\rm cm}-m^b_1-\Sigma_{11}&&-\Sigma_{12}&&-\Sigma_{13}\\
-\Sigma_{21}     &&E_{\rm cm}-m^b_2-\Sigma_{22}&&-\Sigma_{23}\\
-\Sigma_{3 1}&&-\Sigma_{32} &&m_N-\Sigma_{33} 
\end{matrix}
\right) \non
\label{3res}
\end{eqnarray}
where $\Sigma$ is the self-energy:
\begin{eqnarray}
\Sigma_{ij}=\sum_\mu  \gamma^c_{\mu;i}\,G^{}_\mu\,\Gamma^a_{j;\mu} \;.
\end{eqnarray} 

It should be noted that the unitarization of Eq.~(\ref{eq:scat}) can lead to dynamically generated poles also in $T^{\text{NP}}$. 
``Dynamically generated'' thus refers to the appearance of resonances induced through the unitarization of $t$-channel, $u$-channel, and contact terms, but not $s$-channel poles.

As outlined in Ref.~\cite{Ronchen:2012eg} the decomposition of Eq.~(\ref{eq:deco1}) is of numerical advantage since the evaluation of the pole part of the amplitude is much less time-consuming than the evaluation of the non-pole part. It is, thus, possible to apply an effective, nested fitting workflow~\cite{Ronchen:2012eg}. Other than that, we do not attribute any physical meaning to bare masses or couplings, but neither to their dressed counterparts of Eq.~(\ref{eq:dressed})
because the dressing is scheme-dependent and the above decomposition into pole and non-pole part is not unique. See Sec.~4.6 of Ref.~\cite{Ronchen:2012eg} and Ref.~\cite{Doring:2009bi} for an in-depth discussion. The only physically well-defined resonance properties are the pole positions and residues of the full amplitude. 

The inclusion of the $\gamma N$ channel is carried out using the semi-phenomenological approach of Ref.~\cite{Ronchen:2014cna}.
The photoproduction multipole amplitude is given by
\begin{multline}
{\cal M}_{\mu\gamma}(q,E_{\rm cm})=V_{\mu\gamma}(q,E_{\rm cm}) \\+\sum_{\kappa}\int\limits_0^\infty dp\,p^2\,
T_{\mu\kappa}(q,p,E_{\rm cm})G^{}_\kappa(p,E_{\rm cm})V_{\kappa\gamma}(p,E_{\rm cm})\ ,
\label{eq:m2}
\end{multline}
where ${\cal M}$ stands for an electric or magnetic multipole, the index $\gamma$ stands for the initial $\gamma N$ channel and $\mu$ ($\kappa$) denotes the final (intermediate) meson-baryon pair. $T_{\mu\kappa}$ is the hadronic half-off-shell matrix of Eq.~(\ref{eq:scat}) with the off-shell momentum $p$ and the on-shell momentum $q$. In the present study $\mu=\pi N$, $\eta N$ and $K\Lambda$. 

The photoproduction kernel $V_{\mu\gamma}$ is parametrized as 
\be
V_{\mu\gamma}(p,E_{\rm cm})=\alpha^\npo_{\mu\gamma}(p,E_{\rm cm})+\sum_{i} \frac{\gamma^a_{\mu;i}(p)\,\gamma^c_{\gamma;i}(E_{\rm cm})}{E_{\rm cm}-m_i^b} \ ,
\label{eq:vg}
\ee
where $\alpha^\npo$ simulates the coupling of the $\gamma N$ channel to the non-resonant part of the amplitude and the vertex function $\gamma^c_{\gamma;i}$ describes the coupling of the photon to a resonance $i$. The hadronic vertex function $\gamma^a_{\mu;i}$ in Eq.~(\ref{eq:vg}) is exactly the same as in the hadronic scattering potential of Eq.~(\ref{eq:vstuct}) to ensure the cancellation of the poles in Eq.~(\ref{eq:vg}). This formulation also allows to excite background and resonances independently without spoiling Watson's theorem. 

In the present study the photon can excite 
all resonance terms $\gamma^c_{\gamma;i}$; the 
non-pole terms $\alpha^\npo_{\mu\gamma}$ are non-zero in those channels for which we have data at our disposal: $\mu=\pi N$, $\eta N$ and $K\Lambda$. Additionally, we allow the excitation of the $\mu=\pi\Delta$ channel. Without including $\pi\pi N$ data in the analysis it makes no sense to switch on 
$\alpha^\npo_{\rho N,\gamma}$ and $\alpha^\npo_{\sigma N,\gamma}$ because those parameters would be superfluous and highly correlated with 
$\alpha^\npo_{\pi\Delta,\gamma}$.

In analogy to the hadronic case, it is possible to express the photoproduction amplitude in terms of a dressed photon vertex~\cite{Ronchen:2014cna}. Yet, no physical meaning can be assigned to that quantity for the same reasons given before. The well-defined quantities are the residues of photoproduction multipoles at the poles, sometimes referred to as photocouplings~\cite{Ronchen:2014cna}.

The bare photon couplings $\gamma^c_{\gamma;i}$ and $\alpha^\npo$ are approximated by energy-dependent polynomials $P^P$ and $P^\npo$:
\begin{eqnarray}
\alpha^\npo_{\mu\gamma}(p,E_{\rm cm})&=& \frac{ \tilde{\gamma}^a_{\mu}(p)}{\sqrt{m_N}} P^{\text{NP}}_\mu(E_{\rm cm})  \nonumber \\
\gamma^c_{\gamma;i}(E_{\rm cm})&=& \sqrt{m^{}_N} P^{\text P}_i(E_{\rm cm}) \ .
\label{eq:vg_poly}
\end{eqnarray} 
The vertex function $\tilde{\gamma}^a_{\mu}$ is equal to $\gamma^a_{\mu;i}$ but independent of the resonance number $i$. The polynomials $P$ read explicitly:
\ba
P^{\text P}_i(E_{\rm cm})&=&  \sum_{j=1}^{\ell_i}  g^{\text P}_{i,j} \left( \frac{E_{\rm cm} - E_s}{m_N} \right)^j
 e^{-\lambda^\po_{i}(E_{\rm cm}-E_s)} 
\non
P^{\text{NP}}_\mu(E_{\rm cm}) &=&  \sum_{j=0}^{\ell_{\mu}} g^{\text{NP}}_{\mu,j} \left( \frac{E_{\rm cm} - E_s}{m_N}\right)^j
 e^{-\lambda^{\text{NP}}_{\mu}(E_{\rm cm}-E_s)} \ \non .
\label{eq:polys}
\ea
In Eq.~(\ref{eq:polys}), $g^{\po(\npo)}$ and $\lambda^{\po(\npo)}> 0$ are multipole-dependent free parameters that are adjusted in fits to experimental data. The upper limits of the summation $l_i$ and $l_\mu$ are chosen as demanded by the data. In the present study, $l_i$, $l_\mu\le 3$ is sufficient to achieve a good fit result. The expansion point $E_s$ is chosen to be close to the $\pi N$ threshold, i.e. $E_s=1077$~MeV.
 
\subsection{The role of contact terms}
Finding a good description of new data while only including a minimum of new resonances should be the strategy for the determination of the resonance spectrum. 
The contact terms provide the needed flexibility in the fit for this, reducing the need to introduce superfluous $s$-channel poles leading to resonances. 

Yet, the question arises if contact terms can be interpreted in any way. For once, the hadron exchange framework is certainly incomplete and requires such terms. They absorb not only crossed processes of the included hadrons that are absent in the formulation but also contributions beyond the included hadrons. Apart from this, it is difficult to attach any physical meaning to these terms. 

As for the photoproduction mechanism, we also model it with such contact terms because it is known that the tree level diagrams cannot describe multipoles even at very low energies in many cases (see, e.g., Ref.~\cite{Bernard:1994gm}).
For some multipoles the tree level diagrams are much larger than the amplitude at intermediate energies~\cite{SAID} although the unitarization can render them smaller. Comparisons of Born terms and cross sections are also given in Ref.~\cite{Davidson:2001rk}. 
Here we restrain from adding the tree level diagrams explicitly and leave it completely free for the data to decide the form of the multipoles. 

Another role played by the hadronic contact terms is their ability to generate resonances, together with $t$- and $u$-channel hadron exchanges. This means that poles in the complex plane can arise without the explicit inclusion of an $s$-channel pole in the potential. We find several such states in this analysis. From the previous discussion, it is clear that we cannot make any statements on the nature of those states because the contact terms do not originate from a any fundamental interaction. 

Yet, if such a state arises, demanded solely by data, it provides a stronger evidence for the presence of a resonance than if an explicit $s$-channel pole is included by hand (implying automatically a biased decision). 

In summary, the hadron exchange framework is needed to a) provide necessary terms for three-body unitary and b) to approximate nearby left-hand cuts that are important for the analytic structure. Exchanges provide some of the ``background'' but for the quantitative data description with a minimum of resonances, contact terms are needed.
 

\section{Results}
\label{sec:results}


\subsection{Data base}
\label{sec:data}

\begin{table*}
\caption{Data included in the fit. A full list of references to the different experimental publications can be found online~\cite{Juelichmodel:online}. }
\begin{center}
\renewcommand{\arraystretch}{1.9}
\begin {tabular}{l|l|r} 
\hline\hline
Reaction & Observables ($\#$ data points) & $\#$ data p./channel \\ \hline
$\pi N\to\pi N$ & {PWA  GW-SAID WI08 \cite{Workman:2012hx} (ED solution) } & 3,760\\
$\pi^-p\to\eta n$ &{$d\sigma/d\Omega$ (676), $P$ (79) }
& 755\\
$\pi^-p\to K^0 \Lambda$ &{$d\sigma/d\Omega$ (814), $P$ (472), $\beta$ (72) }
& 1,358\\
$\pi^-p\to K^0 \Sigma^0$ &{$d\sigma/d\Omega$ (470), $P$ (120) }
& 590\\
$\pi^-p\to K^+ \Sigma^-$ &{$d\sigma/d\Omega$ (150)}
& 150\\
$\pi^+p\to K^+ \Sigma^+$ &{$d\sigma/d\Omega$ (1124), $P$ (551) , $\beta$ (7)}
& 1,682\\ 
 \hline
$\gamma p\to \pi^0p$ & $d\sigma/d\Omega$ (10743), $\Sigma$ (2927), $P$ (768), $T$ (1404), $\Delta\sigma_{31}$ (140),& \\
& $G$ (393), $H$ (225), $E$ (467), $F$ (397), $C_{x^\prime_\text {L}}$ (74), $C_{z^\prime_\text{L}}$ (26)
&17,564 \\
$\gamma p\to \pi^+n$ & $d\sigma/d\Omega$ (5961), $\Sigma$ (1456), $P$ (265), $T$ (718), $\Delta\sigma_{31}$ (231), &\\
& $G$ (86), $H$ (128), $E$ (903)
& 9,748 \\
$\gamma p\to \eta p$ &$d\sigma/d\Omega$ (5680), $\Sigma$ (403), $P$ (7), $T$ (144), $F$ (144), $E$ (129)
& 6,507\\
$\boldsymbol{\gamma p\to K^+\Lambda}$ & 
$d\sigma/d\Omega$ (2478) \cite{Jude:2013jzs, McCracken:2009ra}, 
$P$ (1612) \cite{McCracken:2009ra, McNabb:2003nf, Lleres:2007tx, 
Glander:2003jw, Tran:1998qw, Bockhorst:1994jf, Haas:1978qv, Fujii:1970gn, Groom:1967zz, Grilli:1965jia, Borgia:1964mza, Thom:1963zz, MD60}, 
$\Sigma$ (459) \cite{Paterson:2016vmc, Hicks:2007zz, Lleres:2007tx, Sumihama:2005er, Zegers:2003ux}, 
 &\\ &
$T$ (383) \cite{Paterson:2016vmc, Lleres:2008em, Althoff:1978qw},
$C_{x^\prime}$ (121) \cite{Bradford:2006ba}, 
$C_{z^\prime}$ (123) \cite{Bradford:2006ba}, 
$O_{x^\prime}$ (66) \cite{Lleres:2008em}, 
$O_{z^\prime}$ (66) \cite{Lleres:2008em} &\\ &
$O_x$ (314) \cite{Paterson:2016vmc}, 
$O_z$ (314) \cite{Paterson:2016vmc}
&5,936 \\ \hline
& \multicolumn{1}{r}{in total} & \bf 48,050\\
\hline\hline
\end {tabular}
\end{center}
\label{tab:data}
\end{table*}

In Tab.~\ref{tab:data} we give an overview of the data analyzed in the current study. 
We include available data for the reactions $\pi N\to\eta N$, $K\Lambda$ and $K\Sigma$ up to $E_{\rm cm}\sim 2.3$~GeV.
For the elastic $\pi N$ channel we fit to the WI08 energy-dependent solution of the GWU/INS SAID partial-wave analysis~\cite{Workman:2012hx}. 
In case of pion and eta photoproduction off the proton the data listed in Tab.~\ref{tab:data} represent the major part of the world database 
up to an energy of 2.3~GeV, including recently published polarization observables such as Refs.~\cite{Akondi:2014ttg,Hartmann:2014mya,
Hartmann:2015kpa,Strauch:2015zob, Senderovich:2015lek, Annand:2016ppc, Thiel:2016chx,Collins:2017sgu}.
Those polarization data were already analyzed in previous J\"uBo fits~\cite{Beck:2016hcy,Strauch:2015zob, Senderovich:2015lek,Collins:2017sgu}, except for the data on $T$ and $F$ for $\gamma p\to \pi^0p$ from Ref.~\cite{Annand:2016ppc}. However, as the prediction from the previous solution J\"uBo2015-B~\cite{Ronchen:2015vfa} is already close to the data, no significant influence on the amplitude or the resonance spectrum is to be expected from the inclusion of this data. Predictions from the J\"uBo2015 solution and the new fit result of the present study are shown in Appendix~\ref{app:tfpi0p}.

For kaon photoproduction, the self-analyzing weak decay of the hyperons facilitates the measurement of the recoil polarization. Accordingly, more data on $P$ but also on the beam-recoil observables $C_{x,z}$ and $O_{x,z}$ are available. Those observables are important to constrain the resonance spectrum and represent a major step towards a complete experiment. 
Recently, the CLAS Collaboration published very accurate data on the polarization observables $\Sigma$, $T$, $O_x$ and $O_z$~\cite{Paterson:2016vmc} which are included in our fit. Note that in Ref.~\cite{Paterson:2016vmc} not only the $K^+\Lambda$ but also the $K^+\Sigma^0$ final state was measured. An analysis of $K\Sigma$ photoproduction within the J\"uBo framework is in progress. 

The data situation for the differential cross section in $\gamma p\to K^+\Lambda$ is ambiguous. While more than 5,500 data points are available in the energy range considered in the present study, not all of them are compatible and systematic discrepancies  beyond that of angle-independent normalization factors can be observed between different experiments.
See, e.g., the discussion of inconsistencies between CLAS and SAPHIR data in Refs.~\cite{Sarantsev:2005tg,Mart:2006dk,Mart:2009nj}. 
We therefore decided to use only the CLAS measurement by McCracken {\it et al.}~\cite{McCracken:2009ra} and the recent MAMI data by Jude {\it et al.}~\cite{Jude:2013jzs} 
and do not include the differential cross sections of Refs.~\cite{Glander:2003jw,McNabb:2003nf,Bradford:2005pt,Sumihama:2005er}. Yet, the comparison of the fit to the entire world data can be found online~\cite{Juelichmodel:online}. The comparison reveals how problematic the data situation is because, e.g., at intermediate energies ($E_{\rm cm}\approx 1.7-1.9$~GeV) some data show a fall-off at extreme forward angles while other continue rising in the forward direction. 
For the polarization observables, on the other hand, no severe inconsistencies occur and all available data are included. 
References to all data considered in the present analysis can be found online~\cite{Juelichmodel:online}.

As can be seen in Tab.~\ref{tab:data} the number of available data points for the different observables and reactions varies considerably. In order to achieve a good fit result for observables with only a few data points, as e.g. $C_x$ in $\gamma p\to \pi^0p$, individual weights are applied in the $\chi^2$ minimization. Otherwise, those data sets would be ignored by the fit. The strategy for choosing the individual weights is to carefully increase the weights for observables with few data points until the description by the fit is improved, provided that the fit result for other data sets is still acceptable. The weights applied in the fit to the $\gamma p\to K^+\Lambda$ data are given in Appendix~\ref{app:weights}.


\subsection{Numerical details}
\label{sec:numerics}

The J\"uBo approach features the following free parameters:
hadronic couplings in the vertex functions $\gamma_{\nu;i}$ and bare masses of the $s$-channel resonances in Eq.~(\ref{eq:vstuct}), coupling constants of the contact diagrams in $V_{\mu\nu}^{\text CT}$ and the parameters connected directly to the photoproduction amplitude, i.e. $g^{\po(\npo)}$ and $\lambda^{\po(\npo)}$ in Eq.~(\ref{eq:polys}). 
Moreover, each $t$- and $u$-channel diagram in $V^\npo$ is multiplied by a form factor and the cut-off parameters in those form factors are treated as free parameters. 
The coupling constants of the exchange diagrams, on the other hand, are related to known couplings via SU(3) flavor symmetry. If this is not possible, the couplings are also fitted to data. See Ref.~\citep{Ronchen:2012eg} for more details.
In the present study, however, we refrain from fitting the parameters tied to $V^\npo$ because the numerical evaluation of the non-resonant part of the scattering matrix, $T^\npo$, is very time-consuming. This is a critical point when fitting several tens of thousands of data as in case of meson photoproduction. Instead we use the values determined in Ref.~\cite{Ronchen:2012eg} in a DCC analysis of pion-induced reactions. Note that we still vary the hadronic contact terms $V_{\mu\nu}^{\text CT}$, i.e., we allow for changes in the hadronic background apart from changes of the hadronic and photonic resonance couplings, and the photoproduction background.

From the 12 $s$-channel poles in isospin $I=1/2$ and 10 $s$-channel poles in $I=3/2$ considered in the present study we get 134 hadronic fit parameters. Those are, for each resonance, one bare mass and the couplings to the channels $\pi N$, $\rho N$, $\eta N$, $\pi\Delta$, $K\Lambda$ and $K\Sigma$ as allowed by isospin.
In addition, we fit the cut-off parameter of the nucleon which is included as an explicit $s$-channel state in the $P_{11}$ partial wave. In contrast, the bare mass and coupling of this state are not free parameters but undergo a renormalization process such that the nucleon pole position and residue to the $\pi N$ channel match the physical values, i.e. $E_0=m_N=938$~MeV and $f_{\pi NN}=0.964$~\cite{Baru:2011bw}. The renormalization procedure is described in the appendix of Ref.~\cite{Ronchen:2015vfa}.
For each partial wave one contact term is included that may couple to the $\pi N$, $\eta N$, $\pi\Delta$, $K\Lambda$ and $K\Sigma$ channel. In practice, the $\pi\Delta$ coupling is only switched on in the $P_{13}$ wave. This amounts to 61 fit parameters from the contact terms.
In case of the fit parameters tied directly to the photoproduction amplitude, $g^{\po(\npo)}_j$ and $\lambda^{\po(\npo)}$, the upper limit of the summation in Eq.~(\ref{eq:polys}) is chosen as demanded by the data but restricted to $j<4$. In the present study we have 566 non-zero parameters.

In total, 761 parameters are adjusted to more than 48,000 points of experimental data in simultaneous fits of all pion- and photon-induced reactions. A systematic reduction of the number of parameters could be performed in the future using model selection techniques~\cite{Landay:2016cjw}. Yet, we consider the large number of free parameters tied to non-resonant contributions an advantage; if that number were too small one would need superfluous resonances making up for missing flexibility of the approach. False-positive resonance signals would be the consequence. We perform a $\chi^2$ minimization using MINUIT on the JURECA supercomputer at the J\"ulich Supercomputing Centre~\citep{jureca}. The code is parallelized in energy; in a typical fit 200-300 processes run in parallel for up to 12 hours.

We estimate the uncertainties of the extracted resonance parameters from re-fits based on re-weighted data sets. To this purpose we individually increase the weight of each of the five pion-induced reactions that are included with experimental data ($\pi^-p\to\eta n$, $K^0\Lambda$, $K^0\Sigma^0$, $K^+\Sigma^-$ and $\pi^+p\to K^+\Sigma^+$) and of the four photon-induced reactions ($\gamma p\to \pi^0p$, $\pi^+n$, $\eta p$, $K^+\Lambda$), imposing that the new $\chi^2$ without a re-fit does not deviate from the best $\chi^2$ by more than 25~\%. After adjusting all free parameters anew the re-fitted $\chi^2$ should not deviate from the best $\chi^2$ by more than 20~\% and the new solution evaluated with the original weights should yield a $\chi^2$ close to the original one. 
The maximal deviations of the resonance parameters of the re-fits from the values of the best fit constitutes the errors quoted in Tabs.~\ref{tab:poles1}, \ref{tab:poles2} and \ref{tab:photo}.
The procedure of re-weighting whole classes of data for a few times, instead of all data individually for, ideally, infinitely many times, is, of course, incomplete and has to be refined when larger computational resources become available.

While this procedure represents only a qualitative estimate of relative uncertainties and the absolute size of the errors is not well determined, it still allows to asses the relative size among the different resonance states. 
A statistically rigorous error analysis including the study of the propagation of statistical and systematic uncertainties from experimental data to partial waves and resonance parameters is beyond the scope of this work; one reason is that the correct inclusion of systematic uncertainties along the lines of the SAID approach. i.e., allowing for angle-independent normalization factors, is not yet fully implemented for all data. Adding systematic and statistical uncertainties in quadrature is not an option because systematic errors are not necessarily Gaussian and, more importantly, induce correlations between data. Even if one allows for multiplicative normalization factors to account for systematic uncertainties, the d'Agostini bias is a problem~\cite{DAgostini:1993arp}, in particular if different experiments are fitted simultaneously~\cite{Ball:2009qv}. 

Moreover, in almost all analysis efforts including the present one, elastic $\pi N$ scattering is not fitted in form of experimental data but in form of partial-wave amplitudes such as the GWU-SAID solution~\citep{Workman:2012hx}. In Ref.~\cite{Doring:2016snk} the covariance matrices and other fit properties of the SAID single-energy solution are provided. This allows to perform correlated $\chi^2$ fits of the SAID partial-wave amplitudes in a statistically meaningful way such that the contribution to the $\chi^2$ from $\pi N$ scattering is very close to the true one that one would obtain from fitting to the data. This method has not yet been included in this analysis. 
Another constraint from elastic $\pi N$ scattering is provided by Roy-Steiner equations~\cite{Hoferichter:2015hva, Hoferichter:2015tha}. Crossing-symmetry and $t$-channel analyticity is used to construct a $\pi N$ amplitude that fulfills these important $S$-matrix principles. In future analyses the low-energy part of the JüBo approach could be matched 
to these new results.

Compared to other analyses, our determination of the uncertainties is somewhere in the middle ground. None of the above problems have ever been fully and consistently addressed, also because of unavoidable weighting factors for certain data sets. As discussed, the SAID approach treats systematic uncertainties better, but pion and photon-induced reactions are never fitted simultaneously as done in this approach. 

Another question is that of model selection and the significance of resonance signals for the determination of a minimal resonance content compatible with data. In that respect, see Ref.~\cite{Landay:2016cjw} where the so-called least absolute  shrinkage and selection operator (LASSO)~\cite{LASSO} was used in an analysis of low-energy pion photoproduction to select the simplest amplitude.  Refs.~\cite{DeCruz:2011xi,DeCruz:2012bv} apply Bayesian evidence to determine the most probable resonance content from kaon photoproduction data. The Bonn-Gatchina group applies mass scan techniques to search for new resonances and in the SAID approach resonances are dynamically generated if required by data. In the present approach we insert only one additional $s$-channel pole by hand, a $N(1900)3/2^+$ that can, of course, change its mass, width and branching ratios in the fit. Additional resonances can still appear through dynamical generation, very similarly as in the SAID approach, if kaon photoproduction data together with all the other data require them.


\subsection{Fit results}

\begin{figure}
\begin{center}
\includegraphics[width=1.\linewidth]{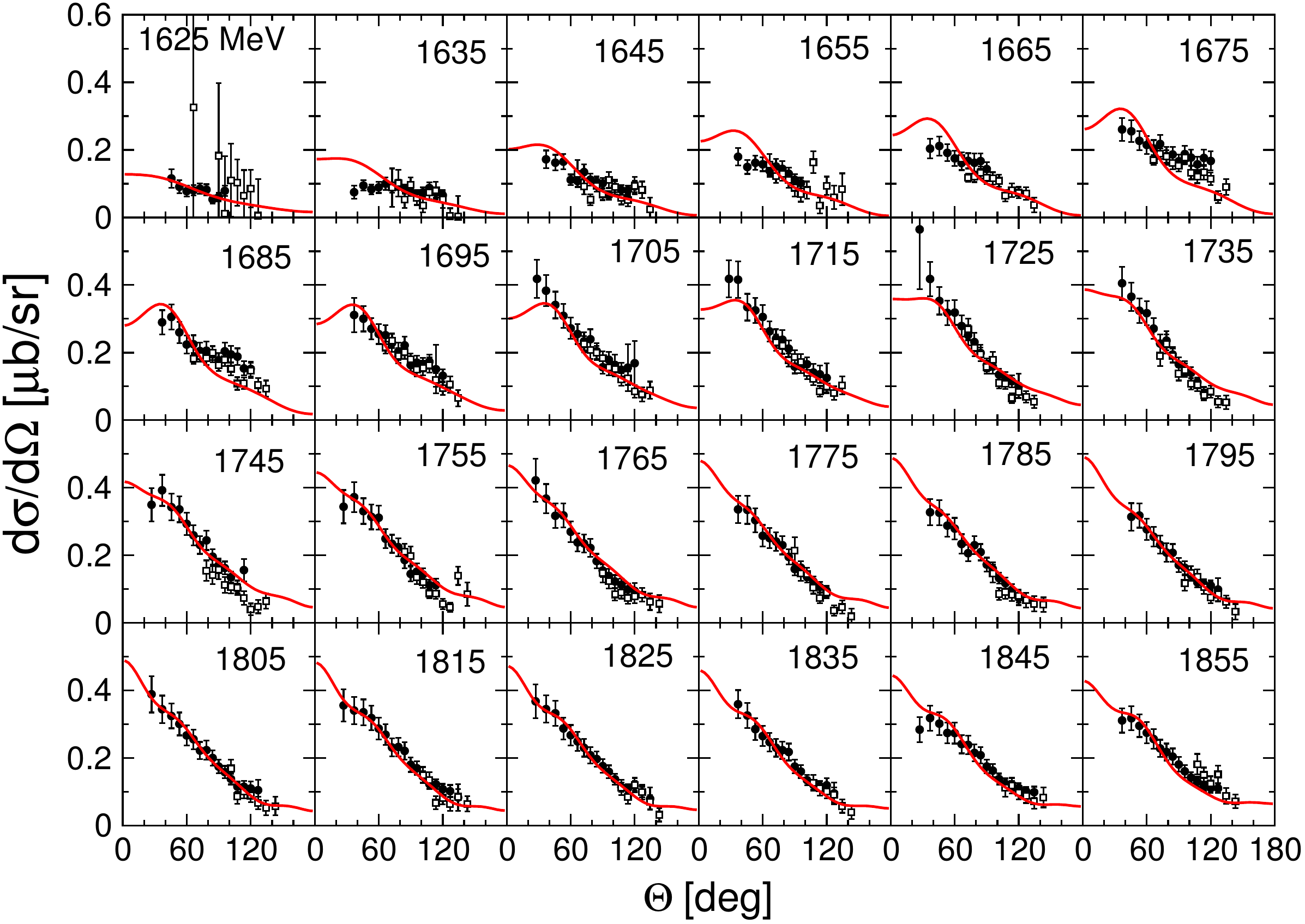} 
\includegraphics[width=1\linewidth]{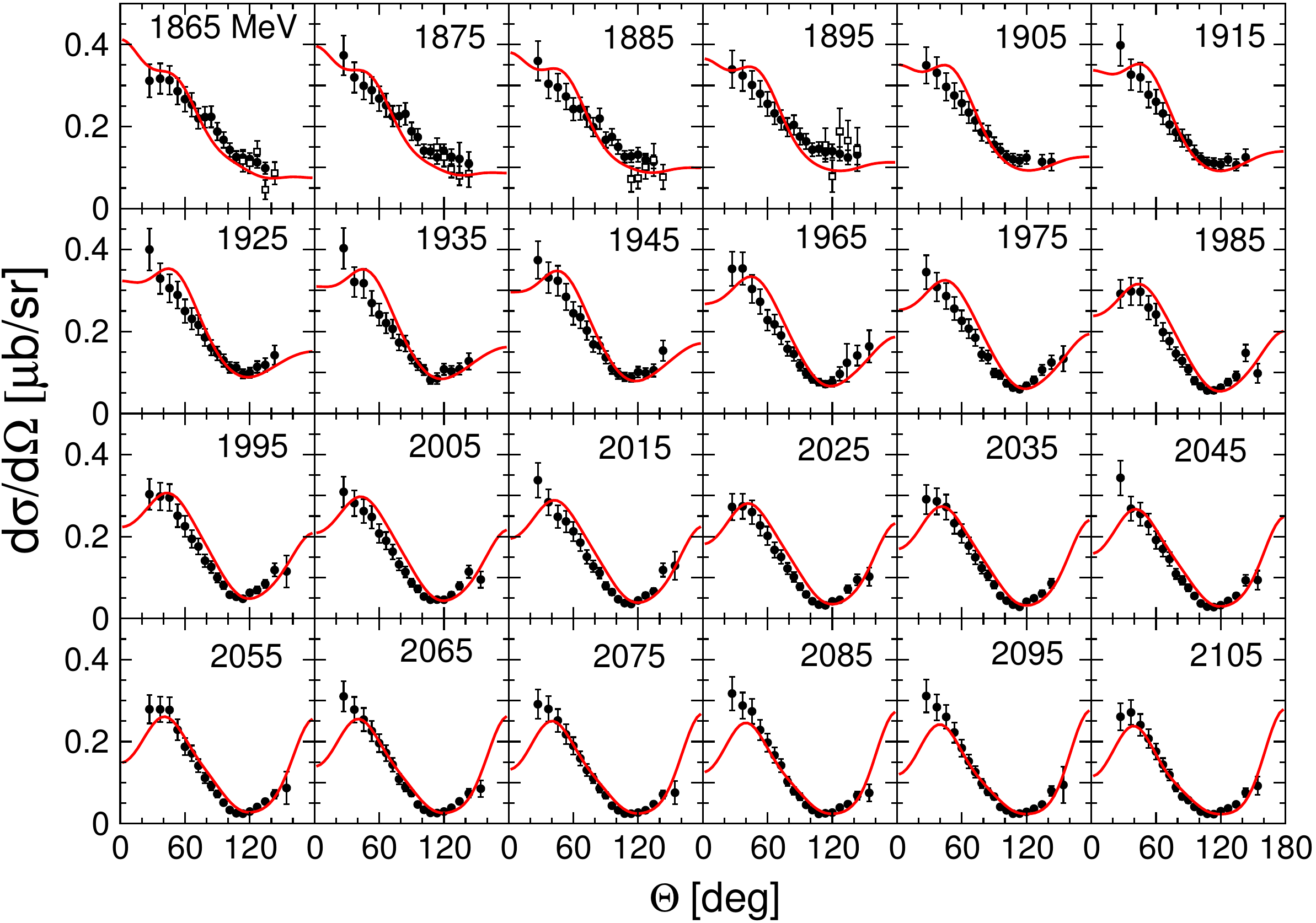} 
\end{center}
\caption{Selected results for the differential cross section of the reaction $\gamma p\to K^+\Lambda$. Filled circles: CLAS (McCracken {\it et al.}~\cite{McCracken:2009ra}); empty squares: MAMI (Jude {\it et al.}~\cite{Jude:2013jzs}). The numbers in the individual panels denote the scattering energy $E_{\text{cm}}$ in the center-of-mass system in MeV.}
\label{fig:dsdoklam1}
\end{figure}

\begin{figure}
\begin{center}
\includegraphics[width=1\linewidth]{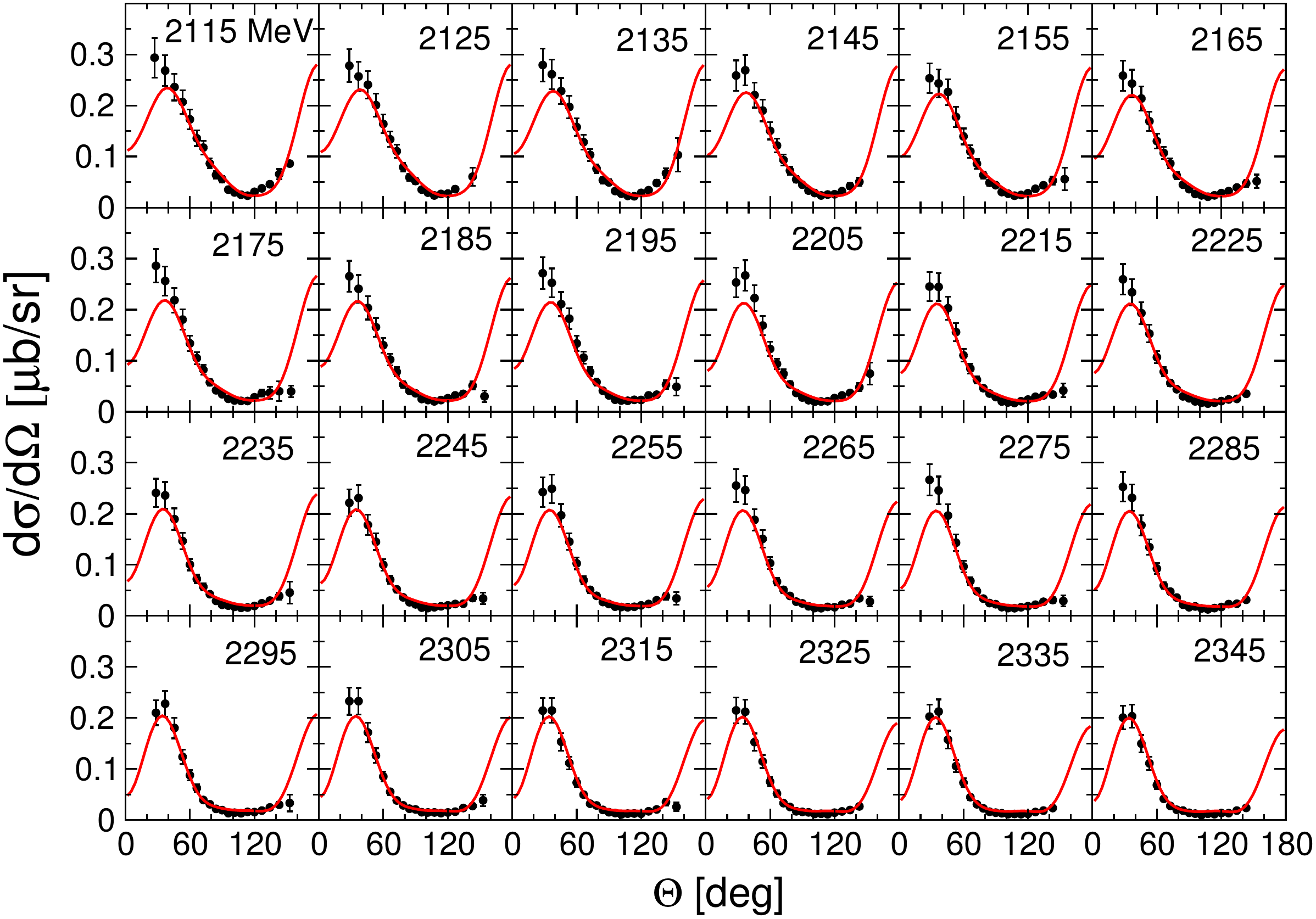} 
\end{center}
\caption{Selected results for the differential cross section of the reaction $\gamma p\to K^+\Lambda$. Filled circles: CLAS (McCracken {\it et al.}~\cite{McCracken:2009ra}). For further notation, see Fig.~\ref{fig:dsdoklam1}.}
\label{fig:dsdoklam2}
\end{figure}

\begin{figure}
\begin{center}
\includegraphics[width=1\linewidth]{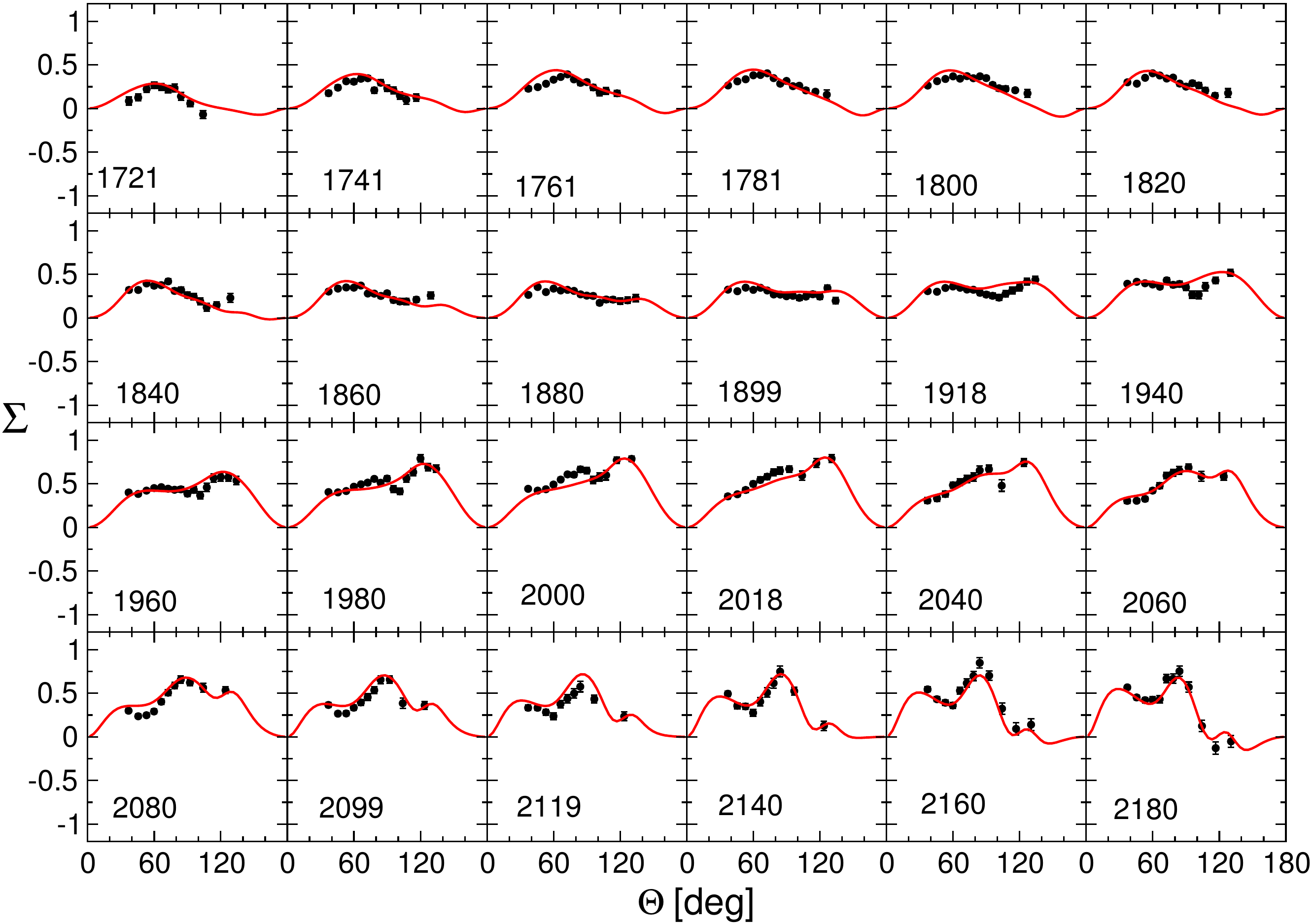} 
\includegraphics[width=1\linewidth]{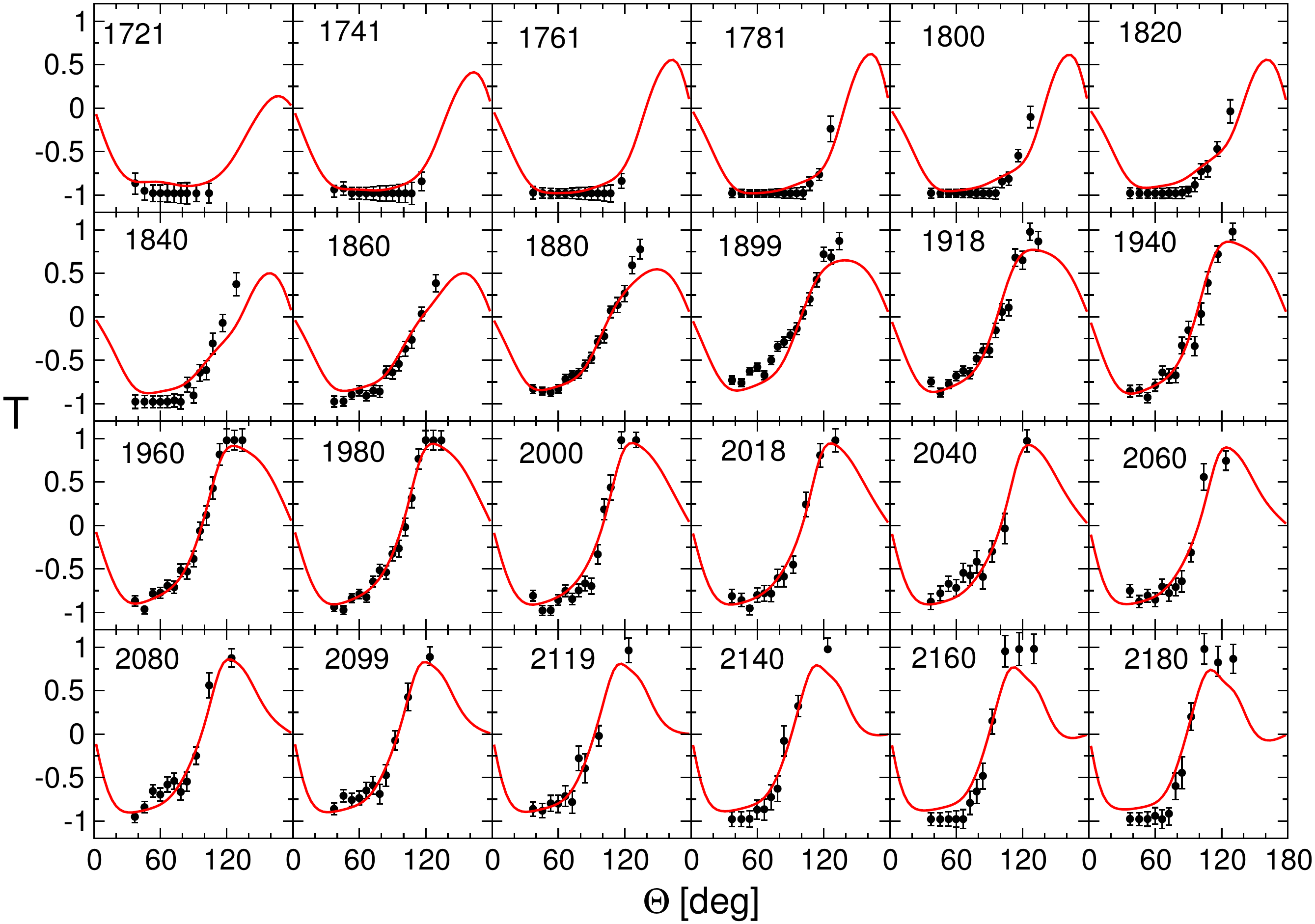} 
\end{center}
\caption{Selected results for the beam asymmetry (above) and target asymmetry (below) of the reaction $\gamma p\to K^+\Lambda$. Data: CLAS (Paterson {\it et al.}~\cite{Paterson:2016vmc}). For further notation, see Fig.~\ref{fig:dsdoklam1}.}
\label{fig:STklam}
\end{figure}

\begin{figure}
\begin{center}
\includegraphics[width=1\linewidth]{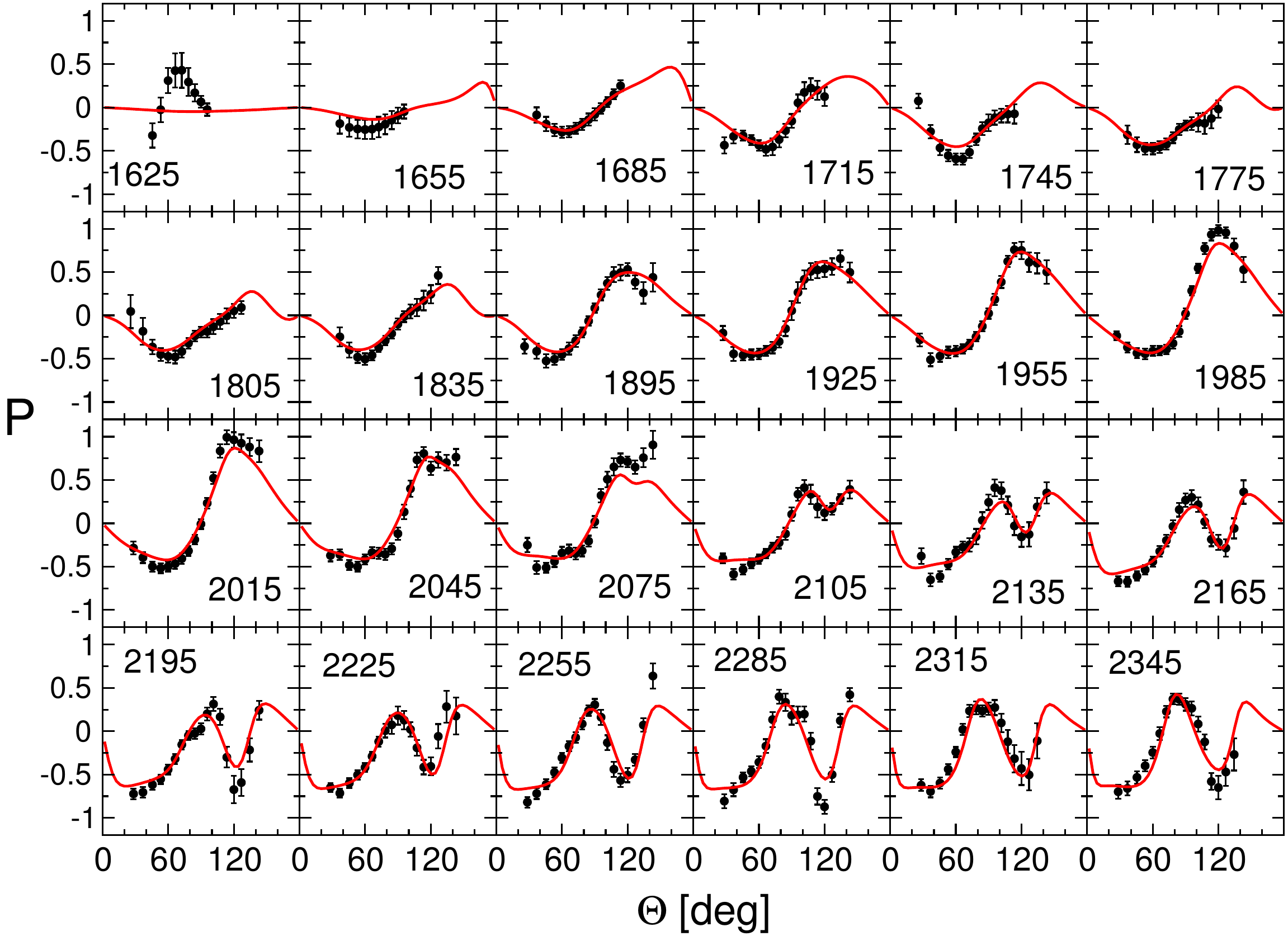} 
\end{center}
\caption{Selected results for the recoil polarization of the reaction $\gamma p\to K^+\Lambda$. Data: CLAS (McCracken {\it et al.}~\cite{McCracken:2009ra}). For further notation, see Fig.~\ref{fig:dsdoklam1}.}
\label{fig:Pklam}
\end{figure}

\begin{figure}
\begin{center}
\includegraphics[width=1\linewidth]{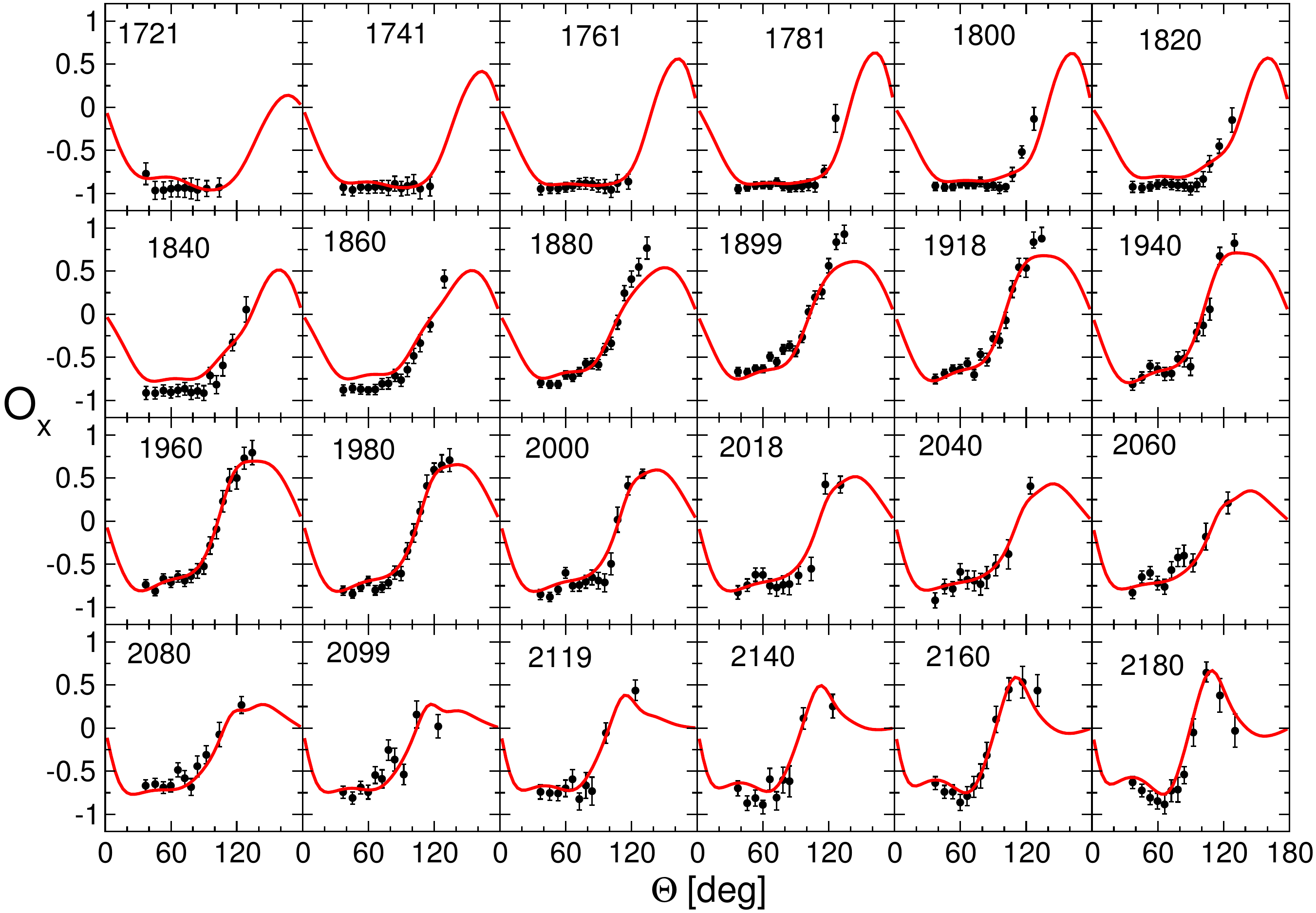} 
\includegraphics[width=1\linewidth]{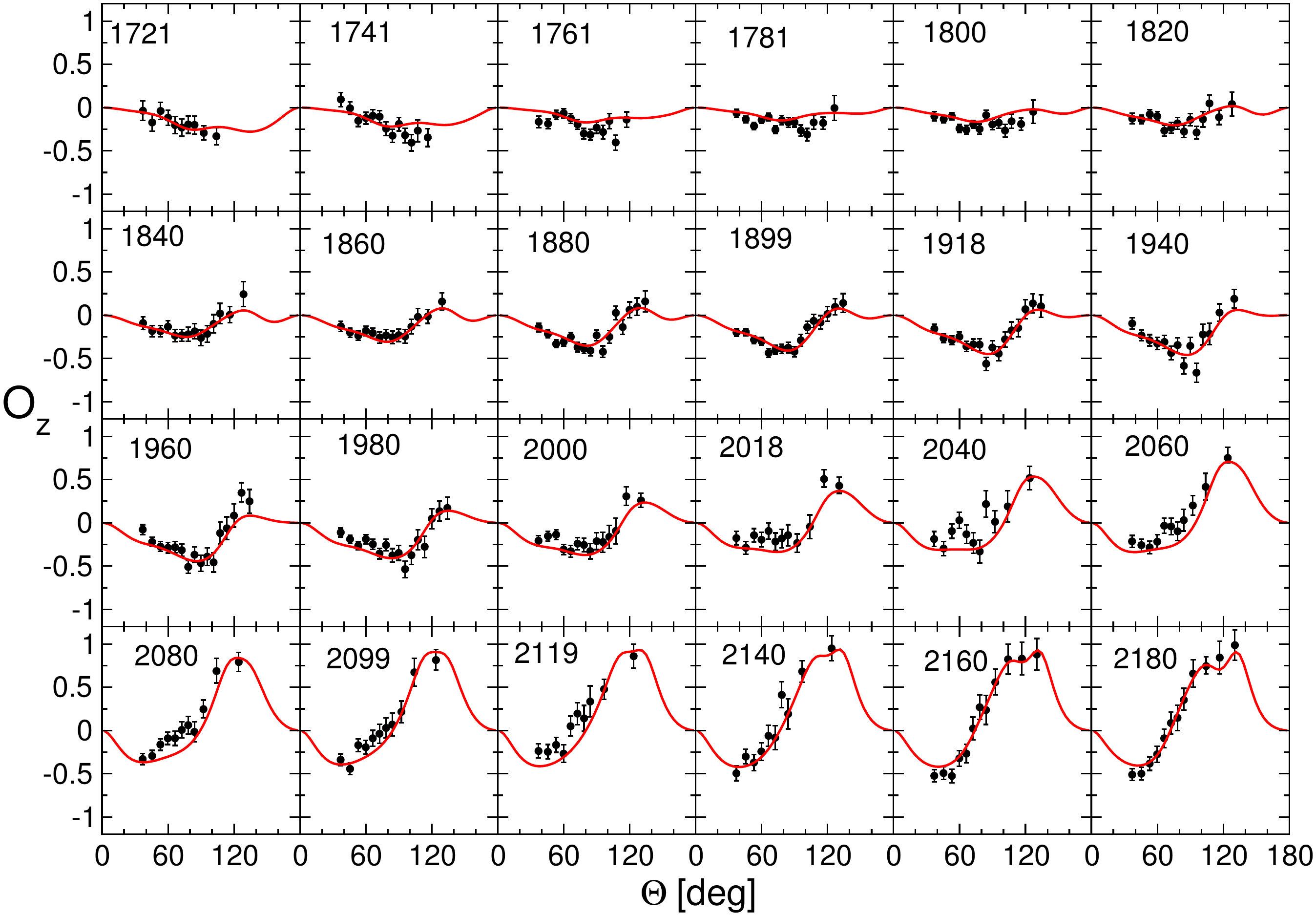} 
\end{center}
\caption{Selected results for the beam-recoil polarization $O_x$ (above) and $O_z$ (below) of the reaction $\gamma p\to K^+\Lambda$. Data: CLAS (Paterson {\it et al.}~\cite{Paterson:2016vmc}). For further notation, see Fig.~\ref{fig:dsdoklam1}.}
\label{fig:OxOzklam}
\end{figure}

\begin{figure}
\begin{center}
\includegraphics[width=1\linewidth]{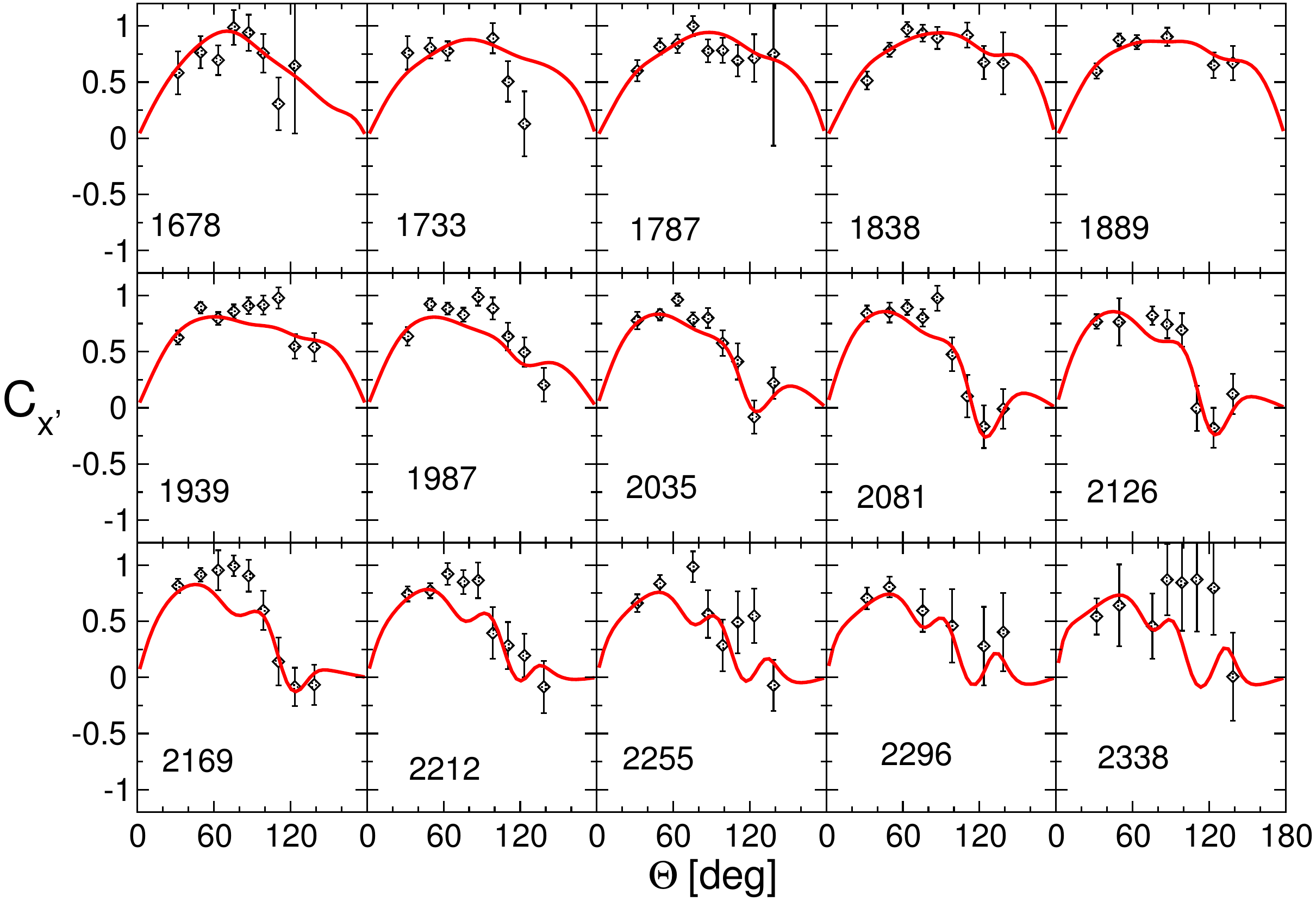} 
\includegraphics[width=1\linewidth]{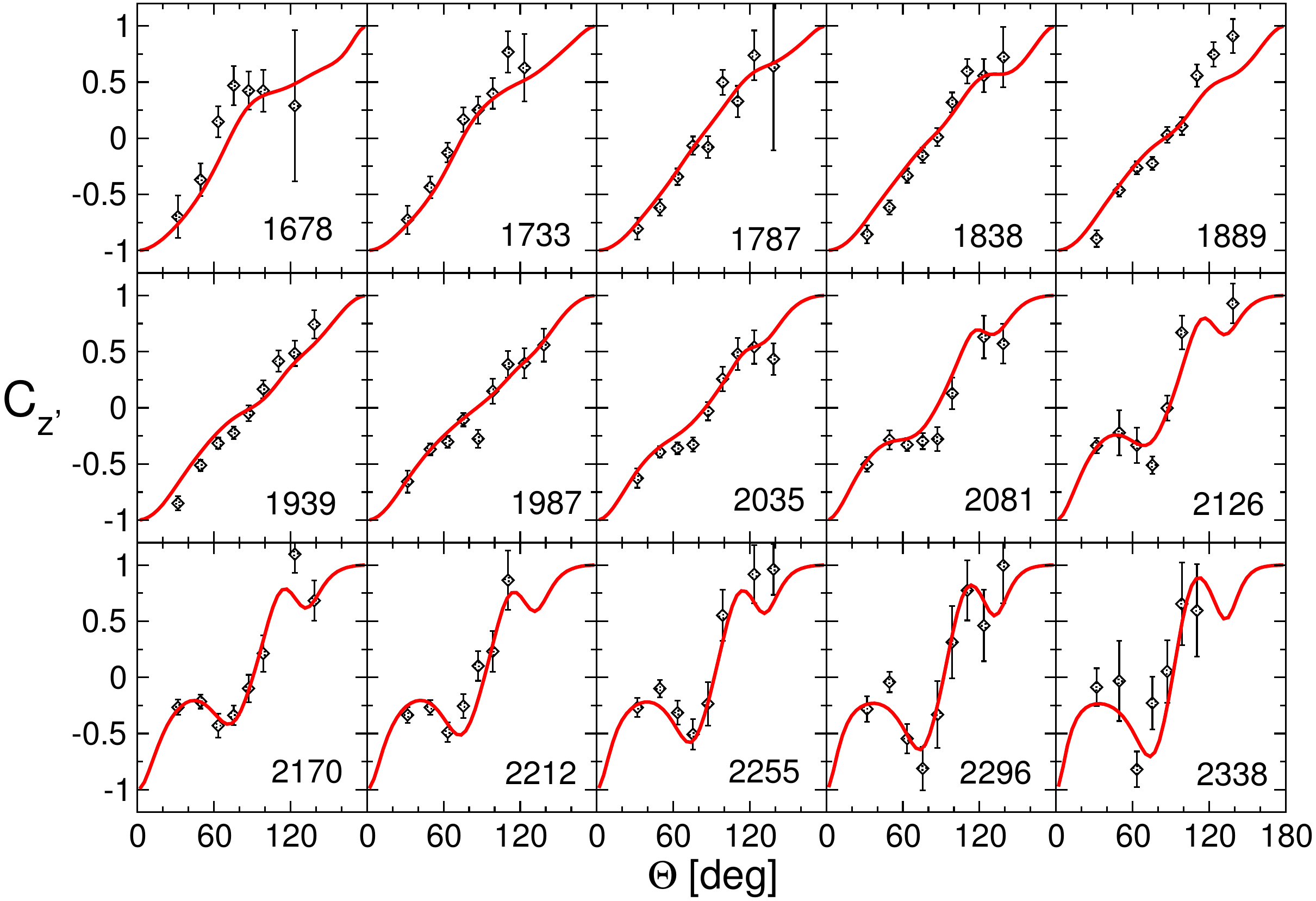} 
\end{center}
\caption{Beam-recoil polarization $C_{x^\prime}$ (above) and $C_{z^\prime}$ (below) of the reaction $\gamma p\to K^+\Lambda$. Data: CLAS (Bradford {\it et al.}~\cite{Bradford:2006ba}). For further notation, see Fig.~\ref{fig:dsdoklam1}. }
\label{fig:CxCzklam}
\end{figure}

\begin{figure}
\begin{center}
\includegraphics[width=1\linewidth]{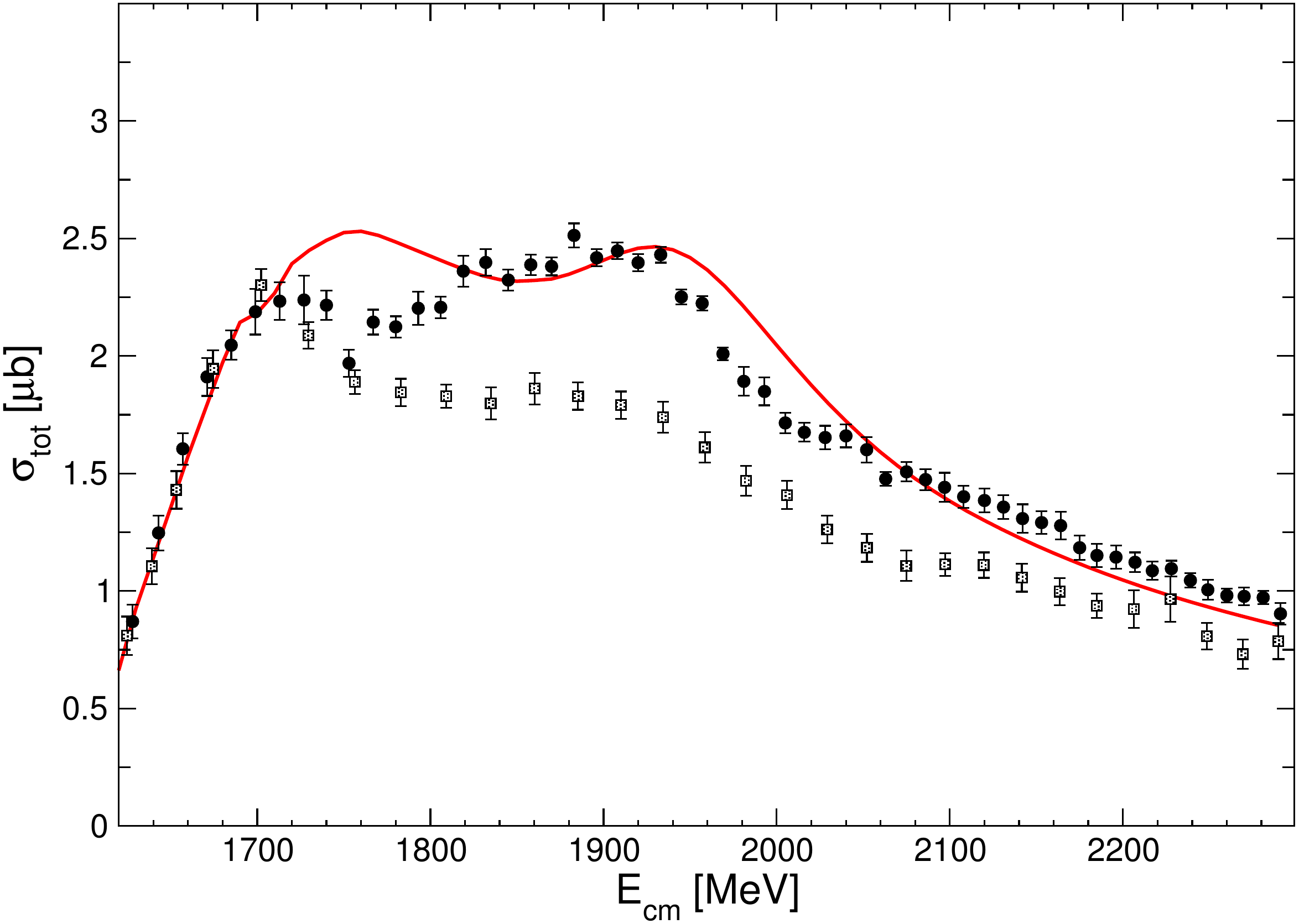} 
\end{center}
\caption{Predicted total cross section of the reaction $\gamma p\to K^+\Lambda$ based on fits to the differential cross sections by McCracken {\it et al.}~\cite{McCracken:2009ra} and Jude {\it et al.}~\cite{Jude:2013jzs}. Data: SAPHIR (Glander {\it et al.}~\cite{Glander:2003jw} (squares)), CLAS (Bradford {\it et al.}~\cite{Bradford:2005pt} (circles)). The data were not included in the fit and imply extrapolations of measurements in the forward direction.}
\label{fig:totcsklam}
\end{figure}

In Figs.~\ref{fig:dsdoklam1} to \ref{fig:CxCzklam} we show selected fit results for the reaction $\gamma p\to K^+\Lambda$. In Fig.~\ref{fig:totcsklam} a comparison of the predicted total cross section with experimental data is shown. Note that these data and also the corresponding differential cross sections were not included in the fit. The discrepancies between the theoretical prediction and the data in Fig.~\ref{fig:totcsklam}, thus, reflect the inconsistencies between the different experimental measurements as the fit result gives a good description of the differential cross section by the CLAS collaboration~\cite{McCracken:2009ra} in Figs.~\ref{fig:dsdoklam1} and \ref{fig:dsdoklam2}. As discussed before, part of these data discrepancies may originate from different extrapolations of the cross section to the forward direction.

The definition of the beam-recoil polarizations $O_x$ and $O_z$ are given in the Appendix, for all other observables the reader is referred to Ref.~\cite{Ronchen:2014cna}.

\begin{figure*}
\begin{center}
\includegraphics[width=1\linewidth]{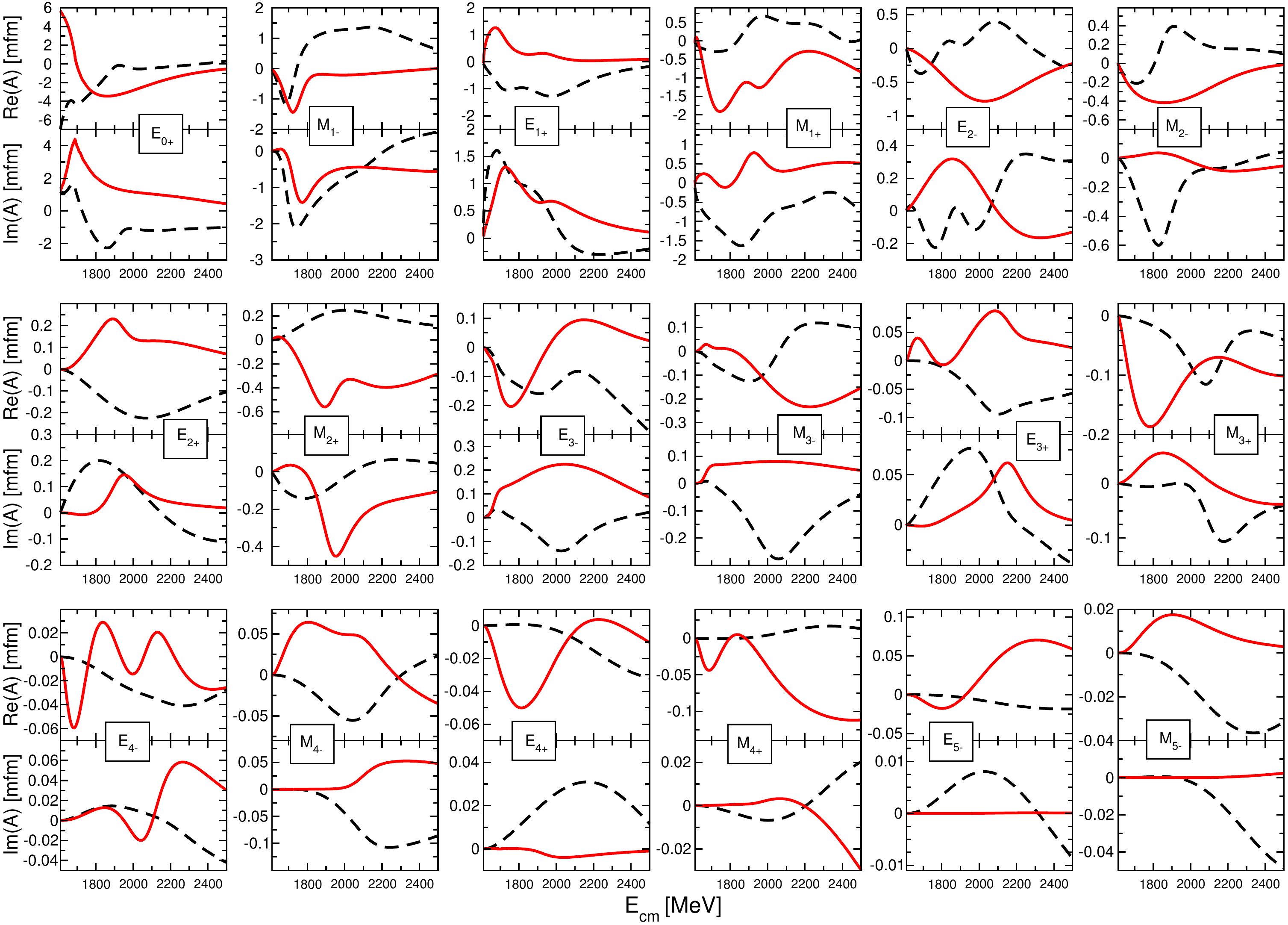} 
\end{center}
\caption{Electric and magnetic multipoles for the reaction $\gamma p\to K^+\Lambda$. (Red) solid lines: J\"uBo2017 (this solution). (Black) dashed lines: BG2014-02 solution~\cite{Gutz:2014wit}. }
\label{fig:mltpklam}
\end{figure*}

Fig.~\ref{fig:mltpklam} represents the $K^+\Lambda$ photoproduction multipoles from the present study and for comparison the BG2014-02 solution from the Bonn-Gatchina partial-wave analysis~\cite{Gutz:2014wit}.
Since there is not yet a complete experiment for $K\Lambda$ photoproduction and, other than in pion or eta photoproduction, no beam-target polarizations are available, significant differences between the two solutions are not surprising. 
In Ref.~\cite{Beck:2016hcy} it was shown for the case of pion photoproduction that multipole amplitudes from different analyses converge indeed to similar solutions if more high-quality polarization data are available. However, it should be noted that additional room for discrepancies arises since the quality of the pion-induced $K\Lambda$ polarization data does not permit an unambiguous determination of the $\pi N\to K\Lambda$ amplitude, which enters as final state interaction into the $K^+\Lambda$ photoproduction process. See, for example, the analyses in Refs.~\cite{Ronchen:2012eg, Shrestha:2012va}. In contrast, in case of pion photoproduction most analysis groups, including Bonn-Gatchina and the present one, use the partial-wave amplitudes from the GW-SAID analysis~\cite{Workman:2012hx} as experimental input for the hadronic final state interaction. It will be more difficult to achieve a unique multipole solution for the $\gamma p\to K^+\Lambda$ process.

Fit results for all other pion- and photon-induced reactions in the present analysis as well as $\pi N$ and $\eta N$ multipoles and $\pi N$ elastic partial-wave amplitudes can be found online~\cite{Juelichmodel:online}.

\setlength{\unitlength}{\textwidth}

\section{Resonance Spectrum}
\label{sec:spectrum}

\subsection{Resonance Parameters}

The resonances are defined as poles in the complex energy plane of unphysical Riemann sheets of the full scattering amplitude. The corresponding residues account for the couplings of the resonances into the different channels. In principle, poles can appear on different sheets except for the physical sheet of the lowest channel, but of physical interest are usually only those on the sheet that is closest to the physical axis. We select this so-called \textit{second sheet} by rotating the right-hand cuts of all channels in the direction of the negative imaginary energy axis. A reliable determination of the resonance parameters requires the correct structure of branch points associated with the opening of inelastic channels. A detailed discussion of the analytic properties of the scattering amplitude can be found in Ref.~\cite{Doring:2009yv}. See also Ref.~\cite{Ceci:2011ae} where the importance of complex branch points related to channels with unstable particles like $\pi\Delta$, $\rho N$ or $\sigma N$ is stressed.

The analytic continuation of the amplitude to the second Riemann sheet is carried out following the method of contour deformation of the momentum integration developed in Ref.~\cite{Doring:2009yv}, and to calculate the residues we apply the formalism illustrated in the appendix of Ref.~\cite{Doring:2010ap}. Definitions of the normalized residue and branching ratios can be found in Ref.~\cite{Ronchen:2012eg}.
The coupling of the $\gamma N$ channel to a given resonance is characterized independently of the hadronic final state by the so-called photocoupling at the pole $\tilde A^h_{pole}$ with
\be
\tilde A^h_{pole}=A^h_{pole} e^{i\vartheta^h}\;.
\label{eq:photocoupling}
\ee
The definition of $\tilde A^h_{pole}$ and its decomposition into electric and magnetic multipoles is given in Appendix C of Ref.~\citep{Ronchen:2014cna}. Note that this definition of the photocoupling agrees with the definition of Ref.~\cite{Workman:2013rca}.

The pole positions and residues for the isospin $I=1/2$ and $3/2$ resonances are given in Tabs.~\ref{tab:poles1} and \ref{tab:poles2}. 
In addition to the values extracted from the fit result of the present study (``J\"uBo2017") we also show the values of the J\"uBo2015 analysis~\cite{Ronchen:2015vfa} for comparison. The latter study included pion and eta photoproduction besides several pion-induced reactions, but not the $\gamma p\to K^+\Lambda$ reaction.

The association of the states found in this analysis with PDG names is clear in most cases but not so clear, e.g., for the $N(2060)5/2^-$ which has a pole more than 100~MeV below the PDG value. Comparing the differences in pole positions, possible uncertainties in the naming of the states should be apparent.
We list the estimated PDG values for pole positions and elastic $\pi N$ residues if available. 
Note that in Tabs.~\ref{tab:poles1}, \ref{tab:poles2} we abbreviate the PDG expression for those estimates due to limited space. E.g., the PDG expression for the real part of the Roper pole position ``1360 to 1385 ($\approx 1370$) OUR ESTIMATE" is converted to the space-saving form of ``$1370^{+15}_{-10}$ ".  
Also note that estimates are only provided by the PDG for resonances rated with three or four stars.
For resonances with less stars we estimate the parameters from the corresponding PDG entries ``above the line'' to have a point of comparison. 
The PDG entries for the ``normalized residues'' all originate from the BnGa group~\cite{Anisovich:2011fc, Sokhoyan:2015fra}.
In Tab.~\ref{tab:poles2} the $\pi\Delta$ channel labeled (6) corresponds to $|J-L|=1/2$ and the one labeled (7) to $|J-L|=3/2$. For easier comparisons, the corresponding orbital angular momenta $L$ are denoted in brackets. See Sec.~4 of Ref.~\cite{Ronchen:2012eg} for details on the angular momentum structure of the coupled channels and a discussion of the meaningfulness of quoting residues or branching ratios for the effective $\pi\pi N$ channels.

In Tab.~\ref{tab:photo} the photocouplings at the pole can be found. Again, we show in addition the values extracted in the J\"uBo2015 analysis to highlight the changes induced by $K\Lambda$ photoproduction.

The uncertainties for all quoted values are in general asymmetrically distributed around the best fit, sometimes the best fit lies even at the border of the uncertainty interval given by the refits specified in Sec.~\ref{sec:numerics}.
We have only performed seven refits to the re-weighted data which can explain this result. 
The uncertainties in the tables, quoted in parenthesis, give the maximal range of the best fit and the seven refits. This is, of course, only a first estimate and to be improved upon in the future. Also, as discussed in Sec.~\ref{sec:numerics} these values indicate relative uncertainties among the resonances. The absolute values are inaccessible to us following the discussion there. In light of this it becomes clear why the uncertainties are in general considerably smaller than in other studies quoted in the PDG~\cite{Patrignani:2016xqp}. This does not suggest higher precision of the current results but indicates that different criteria for the uncertainty determination are used by different groups. Accessing the absolute uncertainties following rigorous statistical criteria remains a challenge for the field.

\begin{table*} \caption{Properties of the $I=1/2$ resonances: Pole positions $E_0$ ($\Gamma_{\rm tot}$ defined as -2Im$E_0$), elastic $\pi N$ residues $(|r_{\pi N}|,\theta_{\pi
N\to\pi
N})$, and the normalized residues $(\sqrt{\Gamma_{\pi N}\Gamma_\mu}/\Gamma_{\rm tot},\theta_{\pi N\to \mu})$ of the inelastic reactions $\pi
N\to \mu$ with $\mu=\eta N$, $K\Lambda$, $K\Sigma$. Resonances with italic numbers in the parentheses are not identified with a PDG state; subscript (a): dynamically generated. We show the results of the present study J\"uBo2017 (``2017") and for comparison the results of fit B of the J\"uBo2015 analysis~\cite{Ronchen:2015vfa} (``2015-B") and the estimates of and from the Particle Data Group~\cite{Patrignani:2016xqp} (``PDG"), if available.
The uncertainties quoted in parentheses provide a rather rough estimate as explained in the text.
 }
\begin{center}
\renewcommand{\arraystretch}{1.3}
\resizebox{\textwidth}{!}{
\begin {tabular}{ll|ll|ll|ll|ll|ll} 
\hline\hline
&&\multicolumn{1}{|l}{Re $E_0$ \hspace*{0.5cm} }
& \multicolumn{1}{l|}{$-$2Im $E_0$\hspace*{0.1cm} }
& \multicolumn{1}{l}{$|r_{\pi N}|$\hspace*{0.2cm}} 
& \multicolumn{1}{l}{$\theta_{\pi N\to\pi N}$ } 
& \multicolumn{1}{|l}{$\displaystyle{\frac{\Gamma^{1/2}_{\pi N}\Gamma^{1/2}_{\eta N}}{\Gamma_{\rm tot}}}$}
& \multicolumn{1}{l|}{$\theta_{\pi N\to\eta N}$\hspace*{0.1cm}}
& \multicolumn{1}{l}{$\displaystyle{\frac{\Gamma^{1/2}_{\pi N}\Gamma^{1/2}_{K\Lambda}}{\Gamma_{\rm tot}}}$} 
& \multicolumn{1}{l|}{$\theta_{\pi N\to K\Lambda}$\hspace*{0.1cm}}
& \multicolumn{1}{l}{$\displaystyle{\frac{\Gamma^{1/2}_{\pi N}\Gamma^{1/2}_{K\Sigma}}{\Gamma_{\rm tot}}}$} 
& \multicolumn{1}{l}{$\theta_{\pi N\to K\Sigma}$}
\bigstrut[t]\\[0.2cm]
&&\multicolumn{1}{|l}{[MeV]} & \multicolumn{1}{l|}{[MeV]} & \multicolumn{1}{l}{[MeV]} & \multicolumn{1}{l}{[deg]} 
& \multicolumn{1}{|l}{[\%]}  & \multicolumn{1}{l|}{[deg]} & \multicolumn{1}{l}{[\%]}  & \multicolumn{1}{l|}{[deg]} &\multicolumn{1}{l}{[\%]} & \multicolumn{1}{l}{[deg]} \\
		  & fit &&&&&&&&
\bigstrut[t]\\
\hline
 $N (1535)$ 1/2$^-$& 2017 &  1495$(2)$  & 112$(1)$ & 23$(1)$ & $-52 (4)$ & 51$(1)$ & 105$(3)$ & 6.0$(1.5)$ & $-44(30)$ & 5.7$(1.6)$ & $-86(6)$   \\
				& 2015-B\hspace{0.2cm} &1499	&104 &22 & $-46$ & 51& 112& 5.0 & 32 & 5.0& $-69$	\\ 
	&PDG& $1510\pm 20$ & $170\pm 80$ & $50\pm 20$& $-15\pm 15$	&$43\pm 3$&$-76\pm 5$&---&---&---&---		\\ \hline
 	 			
 $N (1650)$ 1/2$^-$& 2017& 1674$(3)$ & 130$(9)$ & 29$(4)$ & $-53(7)$ & 18$(3)$ & 28$(5)$ & 17$(1)$ & $-59(3)$ &21$(2)$ & $-67(4)$    \\
 &   	2015-B	& 1672	& 137 & 37 & $-59$ & 21 & 48 & 20 &$-54$ & 26 & $-74$	\\
 &PDG &$1655\pm 15$ &$135\pm 35 $ & $35\pm 15 $ & $70^{+10}_{-20}$ & $29\pm 3$&$134\pm 10$&$23\pm 9$&$85\pm 9$&---&---\\
				\hline
 
 $N (1440)$ 1/2$^+_{(a)}$& 2017& 1353$(4)$ & 213$(2)$ & 62$(2)$ & $-100(7)$ & 8.6$(0.9)$ & $-29(7)$ & 4.8$(0.4)$ & 129$(6)$ & 2.1$(0.4)$ & 87$(22)$ \\
 &   2015-B	& 1355	& 215& 62 & $-98$ & 7.8 & $-27$ & 16 & 145 & 2.7 & 113\\
  &PDG & $1370^{+15}_{-10}$ &$180^{+15}_{-20}$ &$46\pm 6$ & $-90\pm 10$ &---&---&---&---&---&---\\
				\hline


 $N (1710)$  1/2$^+$& 2017& 1731(7) & 157$(6)$ & 1.5$(0.1)$ & 178$(9)$ & 1.6$(0.4)$ & $-137(46)$ & 10$(1)$ & 52$(5)$ & 1.4$(0.1)$ & $-79(24)$ \\
 &   2015-B	&1651	& 121& 3.2 & 55 & 16& $-180$ & 12 &$-32$ & 0.4 &$-43$\\
  &PDG &$1720\pm 50$ & $ 230\pm 150$ &$8^{+7}_{-3}$ &$-$ &$12\pm 4$ &$0\pm 45$ &$17\pm 6$&$-110\pm 20$&---&---\\
				\hline
 
 $N (\textit{1750})$  1/2$^+_{(a)}$& 2017& 1750$(2)$ & 318$(3)$ & 2.9$(2.8)$ & 100$(29)$ &0.7$(0.5)$ & $-31(30)$ & 1.0$(0.2)$ & 164$(19)$ & 3.2$(0.6)$ & 29$(15)$ \\
 &   	2015-B	& 1747 & 323 & 14 & $-144$ & 0.2 & 138&0.4& 86& 1.6& $-55$ \\
				\hline

 $N (1720)$ 3/2$^+$& 2017& 1689$(4)$ & 191$(3)$ & 2.3$(1.5)$ & $-57(22)$ & 0.3$(0.2)$ & 139$(35)$ & 1.5$(0.9)$ & $-66(30)$ & 0.6$(0.4)$ & 26$(58)$ \\
 &   2015-B	& 1710	& 219 & 4.2 &$-47$ & 0.7 & 106 & 1.1 &$-70$ & 0.2 & 79 \\ 
  &PDG &$1675\pm 15$ &$250^{+150}_{-100} $ &$15\pm 8$ & $-130\pm 30$ &$3\pm 2$ &---&$6\pm 4$ &$-150\pm 45$&---&---\\
				\hline
 
 $N (1900)$ 3/2$^+$& 2017& 1923$(2)$ & 217$(23)$ & 1.6$(1.2)$ & $-61(121)$ & 1.1$(0.7)$ & $-10(79)$ & 2.1$(1.4)$ & 1.7$(86)$ & 10$(7)$ & $-34(74)$ \\
  &PDG &$1920\pm 20$ & 130 to 300 &$4\pm 2$ &$-20\pm 40$&$5\pm 2$ &$70\pm 60$ &$7\pm 3$&$135\pm 25$&$4\pm 2$&$110\pm 30$\\
				\hline
 
 $N (1520)$ 3/2$^-$& 2017& 1509$(5)$ & 98$(3)$ & 33$(6)$ & $-16(23)$ & 3.7$(0.6)$ & 85$(18)$ & 0.8$(0.3)$ & 83$(43)$ & 3.0$(1.0)$ &$-28(21)$\\
 	&  2015-B	& 1512	& 89 & 37 & \hspace{-0.cm}$-6$ & 2.6 & 95 & 6.9 & 158 & 4.9 & $-41$\\
 	 &PDG &$1510\pm 5$ &$110^{+10}_{-5}$ &$35\pm 3$ &$-10\pm 5$ &--- &--- &---&---&---&---\\
				 \hline
 
 $N (1675)$ 5/2$^-$& 2017&  1647$(8)$ & 135$(9)$ & 28$(2)$ & $-22(3)$ & 9.1$(1.8)$ & $-45(3)$ & 0.7$(0.2)$ & $-91(6)$ & 2.3$(0.2)$ & $-175(10)$\\
 &  2015-B	& 1646	& 125  & 24 & $-22$ & 4.4 & $-43$ & 0.1 & 100 & 3.1 & $-175$\\
  &PDG &$ 1660\pm 5$ &$135^{+15}_{-10} $ &$27\pm 5$ &$-25\pm 6 $ &--- &--- &---&---&---&---\\
				\hline

 $N ( 2060)$ 5/2$^-_{(a)}$& 2017& 1924$(2)$ & 201$(3)$ & 0.4$(0.1)$ & 172$(12)$ & 0.2$(0.2)$ & 109$(20)$ & 2.2$(0.2)$ & $-86(3)$ & 3.1$(0.3)$ & 86$(3)$   \\
&PDG & $2070\pm 50$ & $385\pm 50$& $22\pm 10$ & $-110\pm 30$& $5\pm 3$ & $40\pm 25$ & $1\pm 0.5$ &--- & $4\pm 2$ & $-70\pm 30$ \\
\hline

 $N (1680)$ 5/2$^+$& 2017& 1666$(4)$ & 81$(2)$ & 29$(1)$ & $-12(1)$ & 1.7$(0.5)$ & 145$(1)$ & 0.9$(0.1)$ & $-77(2)$ & $<0.1$ & $-33(161)$  \\
 &  2015-B	& 1669	& 100 & 34 & $-19$ & 2.7 & 136 & 0.1 & 90 & 0.4 & 148\\ 
  &PDG &$1675^{+5}_{-10}$ &$120^{+15}_{-10}$ &$40\pm 5 $ &$-10\pm 10 $ &--- &--- &---&---&---&---\\
				\hline

 $N (1990)$ 7/2$^+$& 2017& 2152$(12)$ & 225$(20)$ & 0.2$(0)$ & 92$(10)$ & 0.4$(0.2)$ & $-9.1(5.5)$ & 1.4$(0.3)$ & $-13(5)$ & 1.5$(0.3)$ & $-18(6)$  \\
 	&2015-B	& 1738	& 188 & 4.3 & $-70$ & 1.3 & $-82$ & 2.2 & $-111$ & 0.5 & 24\\ 
&PDG &$1965\pm 80$ &$250\pm 60$ &$6\pm 5$ &$30\pm 130$ &--- &--- &---&---&---&---\\
				\hline
 
 $N (2190)$  7/2$^-$& 2017& 2084$(7)$ & 281$(6)$ & 20$(2)$ & $-31(1)$ & 1.2$(0.6)$ & 140$(1)$ & 3.7$(0.3)$ & $-47(1)$ & 0.3$(1.1)$ & 124$(2)$\\
 &   2015-B	& 2074	& 327 & 35 & $-40$ & 1.6 & 129 & 0.5 & $-51$ & 1.3 & $-69$	\\ 
  &PDG &$2075\pm 25 $ &$450^{+70}_{-50} $ &$50^{+20}_{-25}$ &$0^{+30}_{-30}$ &--- &--- &$3\pm 1$&$20\pm 15$&---&---\\
				\hline
 
 $N (2250)$ 9/2$^-$& 2017& 1910$(53)$ & 243$(73)$ & 0.4$(0.1)$ & $-56(25)$ & 0.9$(0.2)$ & $-80(21)$ & $<0.1$ & $-96(21)$ & 0.2$(0.2)$ & $-110(19)$  \\
 &   2015-B	& 2062	& 403 & 8.2 & $-64$ & 1.7 & $-89$ & 0.6 & $-101$ & 0.2 & 70 \\ 
  &PDG &$2200\pm 50 $ & $450\pm 100 $ &$25\pm 5 $ &$-40\pm 20 $ &--- &--- &---&---&---&---\\
				\hline
 
 $N (2220)$ 9/2$^+$& 2017& 2207$(89)$ & 659$(140)$ & 91$(47)$ & $-68(16)$ & 0.3$(0.4)$ & $-109(17)$ & $<0.1$ & 31$(150)$ & 1.0$(0.9)$ & 44$(19)$ \\
 & 2015-B	&2171	&  593 & 62 & $-59$ & 0.4 & $-101$ & 0.7 & 62 & 0.9 & $-128$\\
   &PDG &$2170^{+30}_{-40}$ &$480\pm 80$ &$45\pm 15 $ &$-50^{+15}_{-10} $ &--- &--- &---&---&---&---\\
\hline\hline
\end {tabular}
}
\end{center}
\label{tab:poles1}
\end{table*}

\begin{table*}
\caption{Properties of the $I=3/2$ resonances: Pole positions $E_0$ ($\Gamma_{\rm tot}$ defined as -2Im$E_0$), elastic $\pi N$ residues $(|r_{\pi N}|,\theta_{\pi N\to\pi N})$, and
the
normalized residues  $(\sqrt{\Gamma_{\pi N}\Gamma_\mu}/\Gamma_{\rm tot},\theta_{\pi N\to \mu})$ of the inelastic reactions $\pi N\to K\Sigma$
and $\pi N\to\pi\Delta$ with the number in brackets indicating $L$ of the $\pi\Delta$ state. Subscript (a): dynamically generated. We show the results of the present study J\"uBo2017 (``2017") and for comparison the results of fit B of the J\"uBo2015 analysis~\cite{Ronchen:2015vfa} (``2015-B") and the estimates of and from the Particle Data Group~\cite{Patrignani:2016xqp} (``PDG"), if available.
The uncertainties quoted in parentheses provide a rather rough estimate as explained in the text.
}
\begin{center}
\renewcommand{\arraystretch}{1.3}
\resizebox{\textwidth}{!}{
\begin {tabular}{ll|ll|ll|ll|ll|ll}  \hline\hline
&&\multicolumn{2}{|l}{Pole position}
 &\multicolumn{2}{|l}{$\pi N$ Residue}
 &\multicolumn{2}{|l}{$K\Sigma$ channel} 
 &\multicolumn{2}{|l|}{$\pi\Delta$, channel (6)}
 &\multicolumn{2}{|l}{$\pi\Delta$, channel (7)}
\bigstrut[t]\\[0.1cm]
&&\multicolumn{1}{|l}{Re $E_0$ \hspace*{0.5cm} }
& \multicolumn{1}{l|}{$-$2Im $E_0$\hspace*{0.1cm} }
& \multicolumn{1}{l}{$|r_{\pi N}|$\hspace*{0.cm}} 
& \multicolumn{1}{l}{$\theta_{\pi N\to\pi N}$ } 
& \multicolumn{1}{|l}{$\displaystyle{\frac{\Gamma^{1/2}_{\pi N}\Gamma^{1/2}_{K\Sigma}}{\Gamma_{\rm tot}}}$}
& \multicolumn{1}{l|}{$\theta_{\pi N\to K\Sigma}$\hspace*{0.1cm}}
& \multicolumn{1}{l}{$\displaystyle{\frac{\Gamma^{1/2}_{\pi N}\Gamma^{1/2}_{\pi\Delta}}{\Gamma_{\rm tot}}}$} 
& \multicolumn{1}{l|}{$\theta_{\pi N\to \pi\Delta}$\hspace*{0.1cm}}
& \multicolumn{1}{l}{$\displaystyle{\frac{\Gamma^{1/2}_{\pi N}\Gamma^{1/2}_{\pi\Delta}}{\Gamma_{\rm tot}}}$} 
& \multicolumn{1}{l}{$\theta_{\pi N\to \pi\Delta}$}
\\
&&\multicolumn{1}{|l}{[MeV]} & \multicolumn{1}{l|}{[MeV]} & \multicolumn{1}{l}{[MeV]} & \multicolumn{1}{l}{[deg]} 
& \multicolumn{1}{|l}{[\%]}  & \multicolumn{1}{l|}{[deg]} & \multicolumn{1}{l}{[\%]}  & \multicolumn{1}{l|}{[deg]} &\multicolumn{1}{l}{[\%]} & \multicolumn{1}{l}{[deg]} \\
		  & fit &&&&&&&&
\bigstrut[t]\\
\hline
 $\Delta(1620)$	1/2$^-$&2017& 1601$(4)$ & 66$(7)$ & 16$(3)$ & $-106(3)$ & 31$(6)$ & $-103(2)$ & --- & --- & $57(4)$ {\footnotesize (D)} & 103$(1)$  \\
 &   	2015-B\hspace{0.2cm} &1600	& 65	 & 16 & $-104$ & 22 & $-105$ & --- &--- & 57 & 105\\
& PDG &$1600\pm 10$ &$ 130\pm 10$ & $17^{+3}_{-2}$ &$-100\pm 10$& ---&---&---&---&$42\pm 6$&$-90\pm 20$ \\
				\hline

 $\Delta(1910)$ 1/2$^+$&2017&  1798$(5)$ & 621$(35)$ &  81$(68)$ & $-87(18)$ & 5.1$(2.2)$ & $-96(58)$ & $53(42)$ {\footnotesize (P)} & 126$(15)$ & ---& ---  	 \\ 
 	&  2015-B	&1799	& 648 & 90 & $-83$ & 1.9& $-123$ & 58 & 131 & ---&---	\\
 	& PDG & $1855\pm 25$& $350\pm 150 $ &$30^{+15}_{-10}$&$130\pm 50$&$7\pm 2$&$-110\pm 30$&$24\pm 10$&$85\pm 35$&---&--- \\
				\hline
 	 			

 $\Delta(1232)$ 3/2$^+$	& 2017& 1215$(4)$ & 97$(2)$ & 48$(1)$ & $-40(2)$ &  &&&& \\
 	& 2015-B	&1218	& 91 & 46 & $-36$ & & & &	\\
 	& PDG &$1210\pm1$ &$ 100\pm 2$ &$51\pm 2$&$-46\pm 1$&&&&& \\
				\hline

 $\Delta(1600)$ 3/2$^+_{(a)}$& 2017&  1579$(17)$ & 180$(30)$ & 11$(6)$ & $-162(41)$ & 13$(7)$ & $-21(40)$ & $31(16)$ {\footnotesize (P)} & 37$(40)$ & $0.6(0.9)$ {\footnotesize (F)} & $-56(117)$	 \\
 	&  2015-B	&1552	& 350	& 23 & $-155$ & 13 & $-5.6$ & 31 & 31 & 1.3 & 29\\
 	& PDG &$1510\pm 50 $ &$275\pm 75 $ &$25\pm15 $ &$180\pm30 $&---&---&$15\pm 4$&$30\pm 35$&
$1\pm 0.5$&---\\
				\hline

$\Delta(1920)$	3/2$^+$&2017& 1939$(141)$ & 838$(38)$ & 26$(9)$ & 96$(35)$ & 14$(3)$ & 146$(18)$ & $2.7(1.0)$ {\footnotesize(P)} & 31$(16)$ & $0.6(0.4)$ {\footnotesize (F)} & $-115(86)$   	 \\
	& 2015-B	&1715	&882 	& 38 & \hspace{0.25cm}146 & 17 & $-35$ & 6.9 & 131 & 1.3 & $-115$\\
	& PDG &$1900\pm 50$ &$300\pm 100 $ &$20\pm 6$&$-100\pm 70$&$9\pm 3$&$80\pm 40$&$20\pm 8$&$-105\pm 25$&$37\pm 10$&$-90\pm 20$ \\
				\hline

 $\Delta(1700)$ 3/2$^-$&2017& 1667$(28)$ & 305$(45)$ & 22$(6)$& $-8.6(32.1)$ & 0.7$(1.8)$ & 176$(152)$ & $4.8(2.0)$ {\footnotesize (D)} & 169$(26)$ & $38(14)$ {\footnotesize (S)} & 146$(30)$   \\
 &   2015-B	&1677	& 305 & 24 & $-7.3$ & 1.1 &$ -147$ & 5.4 & 166& 39 & 151	\\
 & PDG &$1650\pm 30 $ &$230\pm 70 $ &$25\pm 15$&$-20\pm20 $&---&---&$12\pm 6$&$-160\pm 30$&$25\pm 12$&$135\pm 45$ \\
				\hline

 $\Delta(1930)$	5/2$^-$&2017& 1663$(43)$ & 263$(76)$ & 5.1$(2.4)$ & $-112(23)$ & 2.5$(0.9)$ & $-27(18)$ & $17(5)$ {\footnotesize (D)} & 68$(17)$ & $0.2(0.2)$ {\footnotesize (G)} & $-134(48)$  \\
 &   2015-B	&1836	& 724 & 34 & $-155$ & 4.3 & $-0.5$ & 15 & 30 & 0.9 & 121	\\
 & PDG &$1900\pm 60 $ &$270^{+90}_{-95} $ &$14\pm 6 $&$-30^{+20}_{-10} $&---&---&---&---&---&--- \\
				\hline

 $\Delta(1905)$	5/2$^+$&2017& 1733$(47)$ & 435$(264)$ & 21$(20)$ & 110$(93)$ & 0.5$(0.5)$ & $-4.3(345)$ & $3.6(3.4)$ {\footnotesize (F)} & $-117(309)$ & $15(15)$ {\footnotesize (P)} & $-61(230)$  	  \\
 	&2015-B	&1795	& 247	& 5.3 & $-89$ & 0.1 & $-155$ & 0.9 & 64 & 8.7 & 72\\
 	& PDG &$ 1820\pm 15$ &$280^{+20}_{-15} $ &$20\pm 5$&$-50^{+20}_{-70} $&---&---&---&---&$19\pm 7$&$10\pm 30$ \\
				 \hline

 $\Delta(1950)$	7/2$^+$&2017& 1850$(37)$ & 259$(61)$ & 34$(20)$ & $-48(46)$ & 1.4$(1.4)$ & $-106(50)$ & $35(25)$ {\footnotesize (F)} & 119$(46)$ & $1.7(1.0)$ {\footnotesize (H)} & $-103(59)$   \\
 	& 2015-B	&1874	& 239 & 56 & $-33$ & 3.1 & $-87$ & 54 & 131 & 3.3 &$ -97$	\\
 	& PDG &$1880\pm 10$ &$240\pm 20 $ &$52\pm 8 $ &$ -32\pm8$ &$5\pm 1$&$-65\pm 25$&$12\pm 4$&---&---&--- \\
				\hline

 $\Delta (2200)$ 7/2$^-$& 2017& 2290$(132)$ & 388$(204)$ & 33$(92)$ & $-32(138)$ & 1.0$(1.0)$ & 118$(165)$ & $7.0(21.1)$ {\footnotesize (G)} & $-103(328)$ & $53(124)$ {\footnotesize (D)} & 137$(132)$ \\
 		& 2015-B	&2142	&486 & 17 & $-56$ &  0.5 & $-103$ & 2.2 & $-151$ & 23 & 107 	\\
 		& PDG &$2100\pm 50$ &$340\pm 80$ &$8\pm 3$&$-70\pm 40$&---&---&---&---&---&--- \\
				 \hline

 $\Delta (2400)$ 9/2$^-$&2017& 1783$(86)$ & 244$(194)$ & 7.2$(8.6)$ & $-78(30)$ & 0.5$(0.6)$ & 9.1$(9.0)$ & $19(9)$ {\footnotesize (G)} & $-95(36)$ & $1.6(1.0)$ {\footnotesize (I)} & $-18(90)$  	\\
 	&2015-B	&1931	& 442 & 13 & $-96$ & 0.9 & 25 & 18 & $-110$ & 1.2 & $-1.0$	\\
 	& PDG &$2120\pm 200$&$600\pm 440$ &$16\pm 12$&$-80\pm 75$&---&---&---&---&---&--- \\
 	 			
\hline\hline
\end {tabular}
}
\end{center}
\label{tab:poles2}
\end{table*}

\begin{table*}
\caption{Properties of the $I=1/2$ (left) and $I=3/2$ resonances (right): 
photocouplings at the pole ($A^h_{pole}$, $\vartheta^h$) according to Eq.~(\ref{eq:photocoupling}). Resonances with italic numbers in the parentheses are not identified with a PDG state; subscript (a): dynamically generated. We show the results of the present study J\"uBo2017 (``2017") and for comparison the results of fit B of the J\"uBo2015 analysis~\cite{Ronchen:2015vfa} (``2015-B"). 
The uncertainties quoted in parentheses provide a rather rough estimate as explained in the text.
}
\begin{center}
\renewcommand{\arraystretch}{1.5}
\resizebox{2.07\columnwidth}{!}{
\begin {tabular}{ll| cc|cc || l l |cc|cc} 
\hline\hline
& & \multicolumn{1}{c}{$\mathbf{A^{1/2}_{pole}}$\hspace*{0.2cm}} 
& \multicolumn{1}{c|}{$\mathbf{\vartheta^{1/2}}$ } 
& \multicolumn{1}{c}{$\mathbf{A^{3/2}_{pole}}$\hspace*{0.2cm}} 
& \multicolumn{1}{c||}{$\mathbf{\vartheta^{3/2}}$ } 
& & & \multicolumn{1}{c}{$\mathbf{A^{1/2}_{pole}}$\hspace*{0.2cm}} 
& \multicolumn{1}{c|}{$\mathbf{\vartheta^{1/2}}$ } 
& \multicolumn{1}{c}{$\mathbf{A^{3/2}_{pole}}$\hspace*{0.2cm}} 
& \multicolumn{1}{c}{$\mathbf{\vartheta^{3/2}}$ } 
\bigstrut[t]\\[0.2cm]
&& \multicolumn{1}{c}{{\footnotesize[$10^{-3}$ GeV$^{-\nicefrac{1}{2}}$]}} &\multicolumn{1}{c|}{\footnotesize[deg]} & \multicolumn{1}{c}{\footnotesize[$10^{-3}$ GeV$^{-\nicefrac{1}{2}}$]} & \multicolumn{1}{c||}{\footnotesize[deg]} 
&&& \multicolumn{1}{c}{\footnotesize[$10^{-3}$ GeV$^{-\nicefrac{1}{2}}$]} &\multicolumn{1}{c|}{\footnotesize[deg]} & \multicolumn{1}{c}{\footnotesize[$10^{-3}$ GeV$^{-\nicefrac{1}{2}}$]} & \multicolumn{1}{c}{\footnotesize[deg]} 
 \\
		  & fit 		&&&&& &fit &	&
\bigstrut[t]\\
\hline
 $N (1535)$ 1/2$^-$  & 2017&106$(3)$  &$-1.6(2.1)$          &&& $\Delta(1620)$	1/2$^-$& 2017&  19$(9)$ &15$(7)$  	\\ 
 &   	2015-B & \hspace{-0.35cm}106 & \hspace{-0.35cm}5.2&&&	&   2015-B & \hspace{-0.32cm}14 & \hspace{-0.32cm}26	\\	\hline
 	 			
 $N (1650)$ 1/2$^-$  & 2017&  51$(3)$ & $-1.4(3.9)$      &&&	$\Delta(1910)$ 1/2$^+$	& 2017&$-238(149)$ &$-87(35)$  	\\
 &   2015-B	& \hspace{-0.32cm}59 & \hspace{-0.32cm}$-14$	&&&&   2015-B & \hspace{-0.32cm}$-321$ & \hspace{-0.32cm}$-141$		\\ \hline
 
 $N (1440)$ 1/2$^+_{(a)}$& 2017&$-90(13)$ &$-33(18)$  &&&  $\Delta(1232)$ 3/2$^+$ & 2017&$ -120(5)$ &$-14(3)$  &$-236(6)$ & 0.5$(1.1)$  	 \\     
 &   2015-B	& \hspace{-0.32cm}$-60$ & \hspace{-0.32cm}$-23$ &&&& 2015-B	&\hspace{-0.32cm}$-117$ & \hspace{-0.32cm}$-6.6$ & \hspace{-0.32cm}$-226$ & \hspace{-0.32cm}2.8 \\
  \hline
 
				 
 $N (1710)$  1/2$^+$  & 2017&$-14(2)$  &$-23(188)$ &&& $\Delta(1600)$ 3/2$^+_{(a)}$& 2017& $-54(25)$ &$-36(31)$   &$-46(19)$  &$-8.5(36)$  \\
 	&   2015-B	& \hspace{-0.32cm}$-20$ & \hspace{-0.32cm}$97$ &&& 			&   2015-B	&\hspace{-0.32cm}$ -230$ & \hspace{-0.32cm}$-42$ & \hspace{-0.32cm}$-332$ &\hspace{-0.32cm}$ 109$		\\
  \hline

  $N (\textit{1750})$  1/2$^+_{(a)}$   & 2017& $-11(7)$ &$11(31)$    &&& $\Delta(1920)$	3/2$^+$& 2017& 35$(15)$  	& $-89(44)$    &77$(17)$  &$-26(39)$ \\
  	&   2015-B	&\hspace{-0.32cm}$-5.0$ &\hspace{-0.32cm}$144$ &&&	&   2015-B	 &\hspace{-0.32cm}$ 192$ & \hspace{-0.32cm}$-134$ & \hspace{-0.32cm}522 & \hspace{-0.32cm}67 \\
   \hline
								
 $N (1720)$ 3/2$^+$& 2017& 48$(24)$&30$(24)$ &$-27(19)$ & $-11(29)$ &  $\Delta(1700)$ 3/2$^-$& 2017& 191$(43)$	& 14$(36)$ &244$(58)$  &$-5.8(32)$ \\ 
 	&   2015-B & \hspace{-0.32cm}39 & \hspace{-0.32cm}5.3 & \hspace{-0.32cm}$-32$ & \hspace{-0.32cm}$-114$	& &   2015-B	 & \hspace{-0.32cm}123 & \hspace{-0.32cm}1.1  & \hspace{-0.32cm}124 & \hspace{-0.32cm}22  	\\\hline

 $N (1900)$ 3/2$^+$& 2017& 34$(13)$ &$-20(65)$  &109$(64)$  &12$(23)$  &  		$\Delta(1930)$	5/2$^-$& 2017& 159$(133)$	&8.7$(26.5)$  & 97$(32)$  &69$(30)$   \\
 &&&&&&& 2015-B & \hspace{-0.32cm}$270$ & \hspace{-0.32cm}$-147$ & \hspace{-0.32cm}153 & \hspace{-0.32cm}81	\\ \hline
 
 $N (1520)$ 3/2$^-$& 2017&$-35(10)$ &$-10(7)$ &77$(17)$ &8.6$(13.1)$  	&  $\Delta(1905)$	5/2$^+$& 2017& 59$(181)$ &11$(235)$  &$-125(295)$ & 28$(195)$    	 \\
 &   2015-B	& \hspace{-0.32cm}$-31$ & \hspace{-0.32cm}$-17$ & \hspace{-0.32cm}75 & \hspace{-0.32cm}1.7 & & 2015-B & \hspace{-0.6cm}53 & \hspace{-0.5cm}89 &  \hspace{-0.65cm}$-30$ &  \hspace{-0.32cm}80	\\  				 \hline
 
 $N (1675)$ 5/2$^-$& 2017&38$(3)$ &17$(10)$  &52$(23)$ &$-11(7)$  &   $\Delta(1950)$	7/2$^+$& 2017&$ -68(29)$ & $-49(35)$ &$-95(43)$ &$-53(46)$  \\
 &   2015-B & \hspace{-0.32cm}32 & \hspace{-0.32cm}36 & \hspace{-0.32cm}51 &  \hspace{-0.32cm}$-9.3$	& 			& 2015-B	&\hspace{-0.32cm}$ -68$ & \hspace{-0.32cm}$-19$ & \hspace{-0.32cm}$-84$ & \hspace{-0.32cm}$-19$ \\ \hline

 $N (2060)$ 5/2$^-_{(a)}$  & 2017& 6.7$(1.6)$  &82$(26)$  &16$(4)$ &50$(12)$  &  $\Delta (2200)$ 7/2$^-$& 2017& 110$(146)$ & 49$(94)$ &57$(69)$  &$-84(64)$  \\
  &&&&&&&    2015-B	& \hspace{-0.32cm}106 & \hspace{-0.32cm}$-23$ & \hspace{-0.32cm}157 &\hspace{-0.32cm}$-60 $	\\  \hline
  		 
 $N (1680)$ 5/2$^+$& 2017&$-8.0(1.8)$&$-42(35)$  &95$(6)$ &$-28(11)$  &   $\Delta (2400)$ 9/2$^-$& 2017&  14$(84)$	&58$(66)$ & 22$(41)$ &89$(82)$  	\\
 	&   2015-B & \hspace{-0.32cm}$-22 $& \hspace{-0.32cm}$-28$ & \hspace{-0.32cm}102 &\hspace{-0.32cm}$ -11$ &				& 2015-B	& \hspace{-0.32cm}$34$ & \hspace{-0.32cm}$-117$ & \hspace{-0.32cm}54 &\hspace{-0.32cm}$ -75$\\  \hline

 $N (1990)$ 7/2$^+$&  2017& $-22(48)$ &13$(236)$  & $-41(69)$ &11$(233)$  & 	  \\ &   2015-B & \hspace{-0.32cm}$-29$ & \hspace{-0.32cm}$-113$ & \hspace{-0.32cm}$-33$ & \hspace{-0.32cm}$-141$	&  		\\ \hline
				 
 $N (2190)$  7/2$^-$& 2017&$-23(13)$ & 70$(40)$  &53$(10)$  &$-82(12)$  & 	 	\\
 	&   2015-B &\hspace{-0.32cm}$-41$ & \hspace{-0.32cm}$-21$ & \hspace{-0.32cm}85 &\hspace{-0.32cm}$-22$	& 			\\ \hline
				 
 $N (2250)$ 9/2$^-$& 2017&$-41(11)$ &$-20(68)$  &20$(15)$  & $-74(60)$& 	\\
 &   2015-B & \hspace{-0.32cm}$-26$ & \hspace{-0.32cm}$154$ & \hspace{-0.32cm}119 & \hspace{-0.32cm}$-42 $&	\\ \hline
				 
 $N (2220)$ 9/2$^+$& 2017& 536$(435)$	& 69$(62)$ & $-445(355)$ & $82(44)$  & \\
 &   2015-B	& \hspace{-0.65cm}135 &\hspace{-0.32cm}114 & \hspace{-0.65cm}$-82$ & \hspace{-0.32cm}$139$ &\\

\hline\hline
\end {tabular}
}
\end{center}
\label{tab:photo}
\end{table*}


\subsection{Discussion of specific resonances }

\begin{figure}[h!]
\begin{center}
\includegraphics[width=1\linewidth]{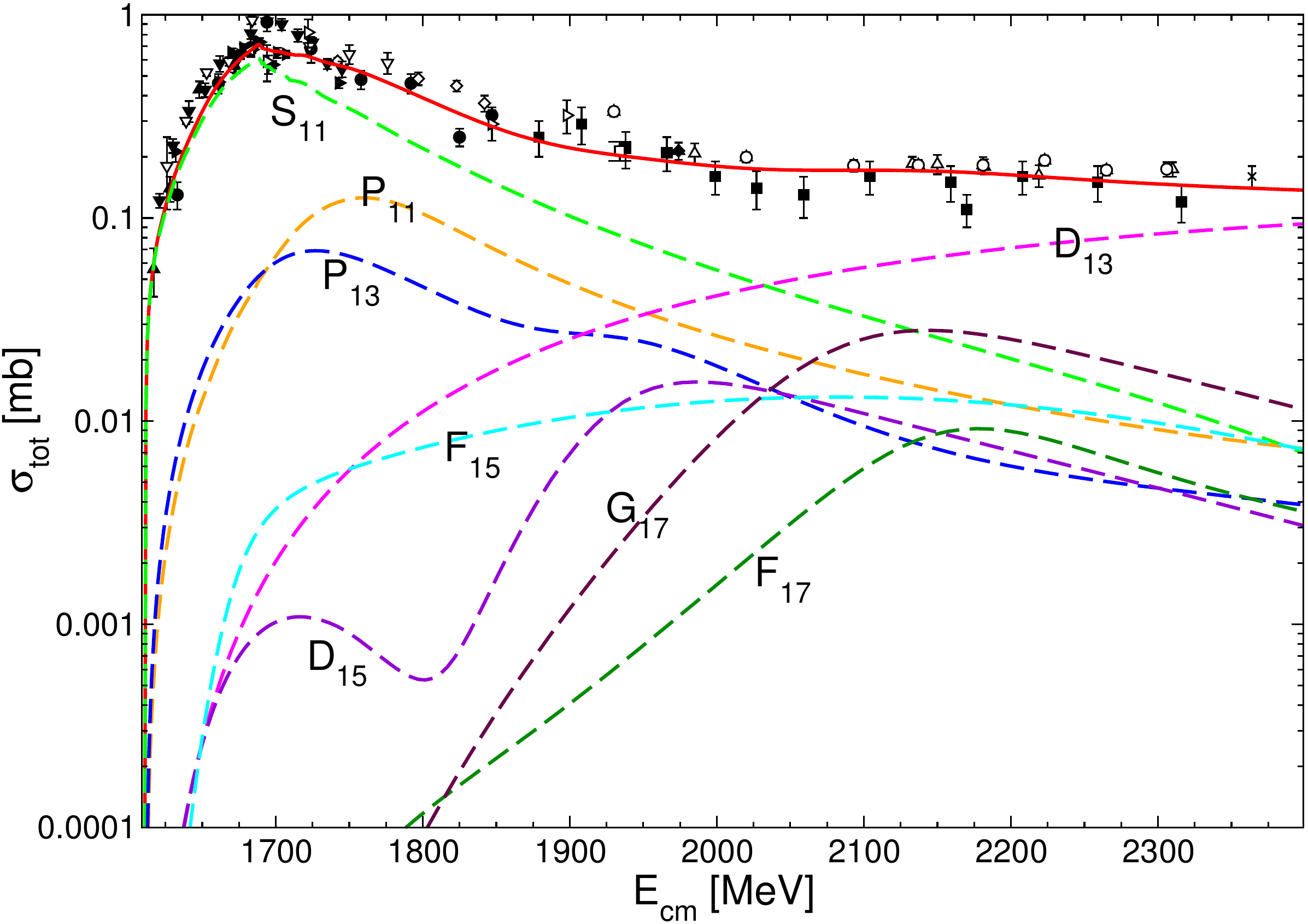} 
\end{center}
\caption{Partial wave content of the total cross section of the reaction $\pi^-p\to K^0\Lambda$ on a logarithmic scale (very small partial waves not shown). Data:  filled
circles: Ref.~\cite{Baker:1978qm}; filled squares: Ref.~\cite{Saxon:1979xu}; empty diamonds: Ref.~\cite{Binford:1969ts}; empty
triangles up: Ref.~\cite{Dahl:1969ap}; filled triangles up: Ref.~\cite{Bertanza:1962pt}; filled triangles down:
Ref.~\cite{Jones:1971zm}; empty triangles down: Ref.~\cite{VanDyck:1969ay}; filled triangles right: Ref.~\cite{Knasel:1975rr}; empty
triangles left: Ref.~\cite{Keren:1964ra}; empty triangles right: Ref.~\cite{Eisler:1958}; empty squares: Ref.~\cite{Yoder:1963zg};
filled diamonds: Ref.~\cite{Goussu:1966ps}; stars:  Ref.~\cite{Miller:1965}; for empty circles and crosses see Ref.~\cite{Landolt}.
}
\label{fig:totcsklamhadropcs}
\end{figure}

\begin{figure}[h!]
\begin{center}
\includegraphics[width=1\linewidth]{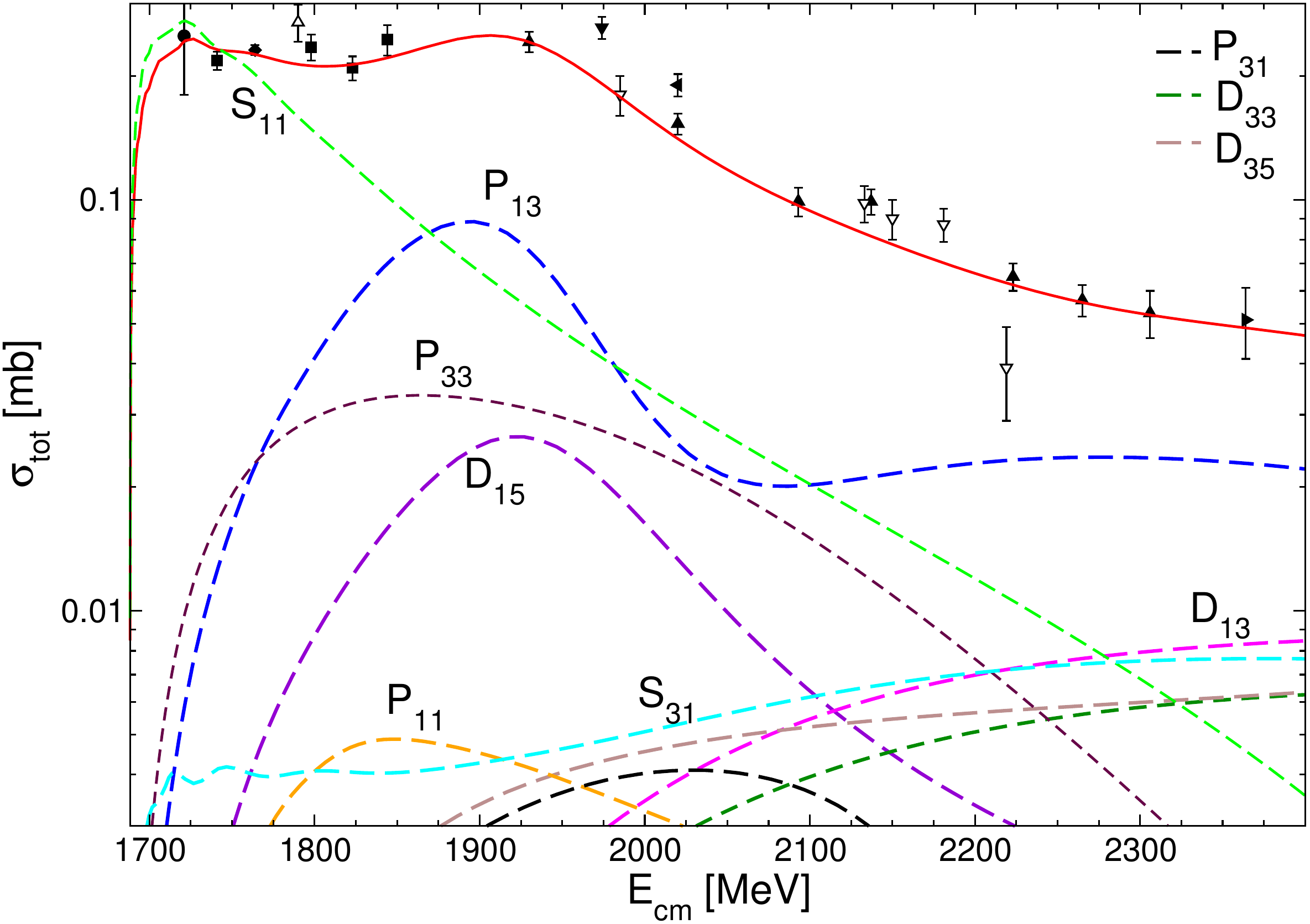} 
\end{center}
\caption{Partial wave content of the total cross section of the reaction $\pi^-p\to K^+\Sigma^-$ on a logarithmic scale (very small partial waves not shown). Data: empty
triangles down: Ref.~\cite{Dahl:1969ap}; filled circles: Ref.~\cite{Eisler:1958}; filled triangles down: Ref.~\cite{Goussu:1966ps}; filled
triangles left: Ref.~\cite{Thomas:1973uh}; empty triangles up: Ref.~\cite{Crawford:1959}; filled squares: Ref.~\cite{Good:1969rb}; filled
diamonds: Ref.~\cite{Doyle:1968zz}. For filled triangles up and filled triangles right see Ref.~\cite{Landolt}. }
\label{fig:totcskpsmhadropcs}
\end{figure}

The 2015 extension of the J\"uBo approach to eta photoproduction did not require the inclusion of additional $s$-channel poles and no new dynamically generated poles were observed. The situation is different in the present study of $K\Lambda$ photoproduction. As will be explained in this section, it was necessary to include a second $s$-channel pole in the $P_{13}$ partial wave and we see evidence of a new dynamically generated resonance in the $D_{15}$ partial wave.

For the discussion of specific resonance parameters we always refer to the values quoted in Tab.~\ref{tab:poles1}, \ref{tab:poles2} and \ref{tab:photo}. In Figs.~\ref{fig:totcsklamhadropcs} and \ref{fig:totcskpsmhadropcs} we show the dominant partial waves in the total cross sections of the reactions $\pi^-p\to K^0\Lambda$ and $K^+\Sigma^-$ that are also relevant for the following discussion.
In the following discussion it should be kept in mind that resonances are defined as poles in the full scattering amplitude, i.e. the resonance states are inevitably superpositions of all components of the $T$-matrix. As the non-resonant hadronic interaction is switched on, bare resonance $s$-channel states in the same partial wave acquire widths and can mix. Yet, sometimes it is possible to identify the mechanism that is primarily responsible for the appearance or the properties of a state. E.g., in case of a resonance associated with a certain $s$-channel diagram, setting some $V^{\rm NP}$ coupling constant to zero might lead to small modifications in the pole position and residues. In contrast, setting the coupling constants of the $s$-channel diagram to zero will lead to the entire disappearance of the pole. 

$\mathbf{S_{11}:}$ The $S_{11}$ partial wave features two resonance state, the well known $N(1535)1/2^-$ and $N(1650)1/2^-$. The pole positions of both states are very similar to the values obtained in previous J\"uBo analyses~\cite{Ronchen:2012eg,Ronchen:2015vfa} and fit in the estimated PDG ranges~\cite{Patrignani:2016xqp}, the real part of the pole position of the  $N(1650)1/2^-$ being only 4~MeV above the estimated upper limit. 
The influence of the $K^+\Lambda$ photoproduction data, not included in earlier J\"uBo studies, on the residues and photocouplings of the $N(1535)1/2^-$ is small, which is no surprise as this state lies below the $K\Lambda$ threshold and is known to couple strongly to $\eta N$. 
In contrast, the couplings of the $N(1650)1/2^-$ into the $KY$ channels are sizeable. Still, compared to earlier J\"uBo studies, which included data on $\pi N\to K\Lambda$, $K\Sigma$, major changes in the extracted values are not observed. 
The normalized residues seem to be already well determined by the data on pion-induced $KY$ production. Also, the photocoupling at the pole is similar to the value extracted in the 2015 analysis of pion and eta photoproduction~\cite{Ronchen:2015vfa}.

$\mathbf{P_{11}:}$ The Roper resonance $N(1440)1/2^+$ is dynamically generated from the interplay of the $t$- and $u$-channel exchanges in the J\"uBo approach. Since the fit parameters of the corresponding diagrams were not altered in the present study and the new $K^+\Lambda$ data enter the fit at energies far above the Roper, the pole position and elastic $\pi N$ residue are very stable when compared to previous J\"uBo studies. They are also close to the values given by the PDG. 

Besides the nucleon whose parameters are renormalized to match the physical values~\cite{Ronchen:2015vfa}, another bare $s$-channel pole is included in the $P_{11}$ partial wave. 
We associate this state with the $N(1710)1/2^+$. 
It was included in the J\"uBo approach for the first time in Ref.~\cite{Ronchen:2012eg} to improve the description of the pion-induced $\eta N$ and $K\Lambda$ channels. In the present study, the coupling to $K\Lambda$ is the dominant one, while the $\eta N$ residue, which was the largest in previous analyses~\cite{Ronchen:2012eg,Ronchen:2015vfa}, is much smaller. 
Moreover, the inclusion of the $\gamma p\to K^+\Lambda$ channel in the present analysis results in a mass 80~MeV higher than in J\"uBo2015. With $E_0=1731-i\,78.7$~MeV the pole position is now within the estimated range of the PDG. We conclude that this state plays an important role in the $K\Lambda$ photoproduction process. This is also reflected in the $M_{1-}$ multipole in Fig.~\ref{fig:mltpklam} where the pronounced dips in the real and the imaginary parts, also found in the Bonn-Gatchina analysis, originate from the $N(1710)1/2^+$.

We find another dynamically generated pole in the $P_{11}$ wave that is not identified with any resonance listed in the PDG. This pole at $E_0=1750-i\,159$~MeV was already seen in previous J\"uBo studies~\cite{Ronchen:2012eg,Ronchen:2015vfa}. The coupling to the $K\Lambda$ channel has increased in this fit compared to the 2015-B solution. Yet, clear evidence is difficult to claim because the pole position is very far in the complex plane and almost behind the pole of the $N(1710)1/2^+$.

A state at around 1900 MeV from three-body interactions has been predicted in this partial wave that could be responsible for the structure seen in kaon photoproduction~\cite{MartinezTorres:2009cw}, see Fig.~\ref{fig:totcsklam}. Here, we not explicitly tested for a new resonance in the $P_{11}$ wave at this energy.

$\mathbf{P_{13}:}$ In the $J^P=3/2^+$ partial wave we include the $N(1720)3/2^+$ as a bare $s$-channel pole. This state was already present in all previous J\"uBo studies. In the present analysis, however, in order to achieve a good fit result in the $\gamma p \to K^+\Lambda$ channel it was necessary to introduce a second $s$-channel diagram. The new pole is associated with the $N(1900)3/2^+$. 
It was observed in several other publications that this resonance plays an important role in the kaon photoproduction process, e.g., by the Bonn-Gatchina Group~\cite{Nikonov:2007br,Anisovich:2017pmi,Anisovich:2017ygb} or in the effective Lagrangian model of Ref.~\cite{Mart:2012fa}. The $N(1900)3/2^+$ is also included in the ANL-Osaka analysis~\cite{Kamano:2013iva} and the Gie\ss en model~\cite{Shklyar:2005xg, Shyam:2009za}. In the present analysis the $N(1900)3/2^+$ has a mass of 1923~MeV and a width of about 217~MeV which is in agreement with the estimated range of the pole position by the PDG. It couples predominantly to $K\Lambda$ and even more to the $K\Sigma$ channel. 
This is reflected in Fig.~\ref{fig:totcskpsmhadropcs} where the pronounced peak in the $\pi^-p\to K^+\Sigma^-$ total cross section at $E_{\rm cm}\sim 1.9$~GeV is induced by the $P_{13}$ partial wave. We expect that the $N(1900)3/2^+$ will play a crucial role also in $K\Sigma$ photoproduction. An analysis within the J\"uBo approach including this channel is in progress. 

The inclusion of the $N(1900)3/2^+$ also results in a change in the pole position of the $N(1720)3/2^+$ which has been rather stable in previous J\"uBo analyses. Compared to the J\"uBo2015-B value of $E_0=1710-i\,109$~MeV the new pole position $E_0=1689-i\,95$~MeV now lies within the estimated range of the PDG.

$\mathbf{D_{13}:}$ In addition to the well established $N(1520)3/2^-$, which couples only weakly to $K\Lambda$, we observe a second pole in the $D_{13}$ partial wave. This  dynamically generated state was already present in the J\"uBo analysis of 2012, where only pion-induced reactions were taken into account. 
It has a mass of $1968$~MeV, i.e. it lies in the energy regime of the 3-star PDG state $N(1875)3/2^-$. Since its width  is very broad, $-2\,\text{Im}\,E_0> 800$~MeV, we do not include this state in our compilation of resonances in Tab.~\ref{tab:poles1}. However, this pole seems to be responsible for the form of the $E_{2-}$ and $M_{2-}$ $K\Lambda$ multipoles in Fig.~\ref{fig:mltpklam}. 
Moreover, as can be seen in Fig.~\ref{fig:totcsklamhadropcs}, $D_{13}$ becomes the dominant partial wave in the total cross section of $\pi^-p\to K^0\Lambda$ at energies $E_{\rm cm}>2$~GeV.
It remains to be seen if further evidence for this state can be obtained in the analysis of $K\Sigma$ photoproduction.
 
$\mathbf{D_{15}:}$ The 4-star $N(1675)5/2^-$ is included as a bare $s$-channel pole. Its parameters are very similar to the ones found in previous J\"uBo studies and close to the PDG values. Since the coupling of this resonance to $K\Lambda$ is comparatively small, major changes in the parameters compared to previous J\"uBo studies do not occur.

In addition to the  $N(1675)5/2^-$ we observe another pole in the $D_{15}$ partial wave at $E_0=1924-i\,100$~MeV that couples predominantly to $K\Lambda$ and $K\Sigma$. This dynamically generated pole was not seen in older J\"uBo calculations~\cite{Ronchen:2012eg,Ronchen:2015vfa} but inconclusive indications for this state were already found in the analysis of the beam asymmetry in the reaction $\gamma p\to \eta p$~\cite{Collins:2017sgu}. Although the parameters of the pole found here differ from the 2 star PDG state $N(2060)5/2^-$ seen in several other analyses, we identify the new pole with the latter state. 
The impact of the $N(2060)5/2^-$ becomes apparent in the $\gamma p\to K^+\Lambda$ multipoles $E_{2+}$ and $M_{2+}$ where the pronounced peak and dip around $E=1900$~MeV in Fig.~\ref{fig:mltpklam} originate from the  $N(2060)5/2^-$  while the $N(1675)5/2^-$ is hardly visible at all. The situation is similar in the cross sections for $\pi^-p\to K^0\Lambda$ and $K^+\Sigma^-$ in Figs.~\ref{fig:totcsklamhadropcs} and \ref{fig:totcskpsmhadropcs}: the $D_{15}$ content exhibits a distinct resonance-like structure at the pole position of the $N(2060)5/2^-$.

$\mathbf{F_{15}:}$ One bare $s$-channel pole is included in this partial wave, the $N(1680)5/2^+$ rated with 4 stars by the PDG. While the real part of the pole position found in the present analyses is in agreement with the PDG value and previous J\"uBo studies, the width is reduced by about 20~MeV. Although the residue of this state into the $K\Lambda$ channel is small, the inclusion of the $\gamma p\to K^+\Lambda$ channel induces this change in the resonance parameters via coupled-channel effects. The changes in the widths are reflected in the photocoupling at the pole which is much smaller in the present analysis than in J\"uBo2015.

$\mathbf{F_{17}:}$ Compared to J\"uBo2015, the present value of the pole position of the $N(1990)7/2^+$ is closer to the value of the Bonn-Gatchina analysis of $E_0=2030\pm 65-i\,(120\pm 30)$~MeV~\cite{Anisovich:2011fc}. 
In contrast to our observation in previous analyses, where we concluded that is was hard to determine the properties of the $N(1990)7/2^+$~\cite{Ronchen:2012eg,Ronchen:2015vfa}, the current data base with $K\Lambda$ photoproduction helps to fix its pole position with smaller relative uncertainty. The photocouplings at the pole, however, still show larger variations.

$\mathbf{G_{17}:}$ The $N(2190)7/2^-$ is included as a bare $s$-channel pole. It couples predominantly to the $\pi N$ channel and has a mass of 2084~MeV which is close to the PDG value and comparable to values found in previous J\"uBo analyses. The width, on the other hand, is reduced in the present fit: $-2\,\text{Im}\,E_0=281$~MeV compared to about 327~MeV in J\"uBo2015 and 450~MeV the estimate of the PDG. This also results in much smaller photocouplings $A^{1/2}$ and $A^{3/2}$.

$\mathbf{G_{19},\,H_{19}:}$ One $s$-channel pole is included in both the $G_{19}$ and the $H_{19}$ partial waves. We identify those states with the $N(2250)9/2^-$ and the $N(2220)9/2^+$. Both resonances exhibit large uncertainties in their parameters and couple very weakly to $K\Lambda$. Still, in case of the $N(2250)9/2^-$ the inclusion of the $\gamma p\to K^+\Lambda$ channel leads to a lower and considerably narrower pole position compared to J\"uBo2015.
The $N(2220)9/2^+$ has a large elastic $\pi N$ residue and is very broad. In view of the fluctuations that are typical for higher lying, broad resonances, the parameters of the $N(2220)9/2^+$ found in the present study are comparable to earlier J\"uBo solutions.

\textbf{Isospin I=3/2 resonances:} Since the $K\Lambda$ final state couples only to resonances with isospin $I=1/2$ one might expect that the inclusion of the $\gamma p\to K^+\Lambda$ channel would not induce major changes in the spectrum of $\Delta$ resonances. On the other hand, in the present analysis the whole data base was refitted, including the mixed-isospin channels with $\pi N$ and $K\Sigma$ final states. Accordingly, adjustments of the $I=1/2$ resonances required to describe the $\gamma p\to K^+\Lambda$ channel will result in changes in the parameters of $\Delta$ states in order to maintain a good description of the pion- and photon-induced $\pi N$ channel and the pion-induced $K^+\Sigma^-$ and $K^0\Sigma^0$ channels.

Most of the well established $\Delta$ resonances are  similar to previous J\"uBo results. Still, we observe, in general, larger uncertainties than in case of the $I=1/2$ states. This is based on the fact that a large part of the current data base stems from reactions that do not couple to isospin $I=3/2$, i.e. reactions with $\eta N$ and $K\Lambda$ final states. We expect that the uncertainties will be reduced once the analysis is extended to $K\Sigma$ photoproduction. 
										
One of the most striking differences to earlier results is the width of the $\Delta(1600)3/2^+$ which is reduced by almost a factor of two compared to the J\"uBo2015-B result. This applies also to the elastic $\pi N$ residue.
The modulus of the photocoupling at the pole $A^{1/2}_{\text{pole}}$ is more than four times smaller, $A^{3/2}_{\text{pole}}$ even more than seven times. The photocouplings are now in good agreement with the values found in a recent Bonn-Gatchina analysis~\cite{Sokhoyan:2015fra}. As in the J\"uBo analysis of Ref.~\cite{Ronchen:2012eg}, this state is dynamically generated with a strong $\pi\Delta$ $P$-wave residue. Changes in other components in the $P_{33}$ wave, i.e. the bare $s$-channel states and the inclusion of a contact term, lead to the changes in the pole position of the $\Delta(1600)3/2^+$ compared to the values found in previous studies.

The $\Delta(1232)3/2^+$ changes its pole position slightly by $(3-i\,2.5)$~MeV and moves closer to the PDG values. The third pole in the $P_{33}$ partial wave, the $\Delta(1920)3/2^+$, is very broad and shows large uncertainties which are typical for a state this far from the physical axis.

In the $J=5/2$ partial waves an interplay between the $\Delta(1930)5/2^-$ and the $\Delta(1905)5/2^+$ seems to occur: the former state was very broad in fit B of the J\"uBo2015 solution, while it is now much narrower. The $\Delta(1905)5/2^+$, on the other hand, is much broader now and was narrower in J\"uBo2015.  Noteworthy is also the large uncertainty in the width of the $\Delta(1905)5/2^+$. As a consequence, also the photocouplings at the pole show large variations since the pole position enters the definition of $\tilde A^h_{pole}$ in Eq.~(\ref{eq:photocoupling}). As the maximal value of the width in the determination of uncertainties was extracted from a re-fit with increased weight on the $\pi^+p\to K^+\Sigma^+$ reaction (see Sec.~\ref{sec:numerics} for methodology), we expect that the extension of the analysis to $K\Sigma$ photoproduction will help to fix the parameters of this state.

The resonances with higher spin are often less stable. Nonetheless, the changes in the pole position of the $\Delta(2400)9/2^-$ are worth mentioning: $E_0=1783-i\,122$~MeV compared to $E_0=1931-i\,221$~MeV in fit B of the J\"uBo2015 solution.
 
In addition to the states listed in Tab.~\ref{tab:poles2} we see indications of a dynamically generated pole in the $P_{31}$ partial wave at Re$\,E_0\sim 2200$~MeV.
We expect that the inclusion of the $K\Sigma$ photoproduction channels in  future analyses will give more information on the $\Delta$ resonance spectrum.


\section{Conclusion}

Kaon photoproduction promises to shed light on the so-called ``missing resonance problem" and reveal resonances that are not observed in non-strange channels as, e.g., $\gamma N\to \pi N$ or $\eta N$. Moreover, the self-analyzing decay of the hyperons facilitates the measurement of recoil polarization observables which are indispensable for a complete set of observables. To extract the baryon spectrum, coupled-channel approaches provide an especially suited tool as they combine several reactions with different initial and final states in a simultaneous analysis.

In the present study, the J\"ulich-Bonn dynamical coupled-channel approach was extended to $K\Lambda$ photoproduction and includes now the photon-induced reactions $\gamma p\to \pi^0p$, $\pi^+n$, $\eta p$ and $K^+\Lambda$ in addition to the pion-induced reactions $\pi N\to \pi N$, $\pi^-p\to\eta n$, $K^0\Lambda$, $K^0\Sigma^0$, $K^+\Sigma^-$ and $\pi^+p\to K^+\Sigma^+$. 
More than 40,000 data points from differential cross sections, single and double polarization observables up to $E_{\rm cm}\sim 2.3$~GeV were analyzed in simultaneous fits to all reactions, and the spectrum of nucleon and $\Delta$ resonances in terms of pole positions, residues and photocouplings at the pole was determined. 

We find all states rated with 4 stars by the PDG and most of the 3-star states and compare our results to earlier J\"ulich-Bonn studies and the estimates of the PDG. 
While most of the well-established resonances are similar to previous studies, we observe noticeable changes in the pole positions of the $N(1710)1/2^+$ and $N(1720)3/2^+$, which move closer to the PDG values.
Moreover, the extension to kaon photoproduction required one additional $s$-channel resonance, the $N(1900)3/2^+$, that was not needed to achieve a good fit result in pion or eta photoproduction. The mass and the width found here are in good agreement with the PDG values. In addition, we observe a new dynamically generated pole in the $D_{15}$ partial wave with significant couplings to the $K\Lambda$ and $K\Sigma$ channels and see indications of further dynamically generated states in the $D_{13}$ and $P_{31}$ wave.

Uncertainties of the resonance parameters were estimated from several re-fits to re-weighted data sets. The pole positions of the nucleon resonances show only small variations with the exception of the broad $J=9/2$ states, while the $\Delta$ resonances are in general less stable. We expect that the latter observation will change once the analysis is extended to the mixed-isospin $K\Sigma$ photoproduction channels.

In summary, the central findings of this study are the confirmation of the $N(1900)3/2^+$ and $N(2060)5/2^-$ resonances although the latter state appears with a pole position significantly different from the PDG value. In addition, many resonances move closer to their PDG values and hints for new states were found. To establish these states $K\Sigma$ photoproduction will be analyzed in the future.

 
\acknowledgments
We thank J. Haidenbauer, K. Nakayama, D. Ireland, T. Jude, R. Schumacher, I. Strakovsky, U. Thoma, D. Watts, and R. Workman for useful discussion and for providing data.
The authors gratefully acknowledge the computing time granted by the JARA-HPC Vergabegremium on the supercomputer JURECA at Forschungszentrum J\"ulich. This work is supported in part by DFG and NSFC through funds provided to the Sino-German CRC 110 ``Symmetry and the Emergence of Structure in QCD" (NSFC Grant No. 11621131001, DFG Grant No. TRR110), and by VolkswagenStiftung (grant No. 93562). M.D. is supported
by the National Science Foundation (CAREER grant
PHY-1452055, NSF/PIF grant No. 1415459) and by
the U.S. Department of Energy, Office of Science, Office
of Nuclear Physics under contract DE-AC05-06OR23177 and under
grant No. DE-SC0016582. 
The work of UGM was also supported by the CAS President's International Fellowship Initiative (PIFI) (Grant No. 2018DM0034).

\appendix

\section{Polarization $T$ and $F$ in $\gamma p\to \pi^0p$}
\label{app:tfpi0p}

\begin{figure}
\begin{center}
\includegraphics[width=1\linewidth]{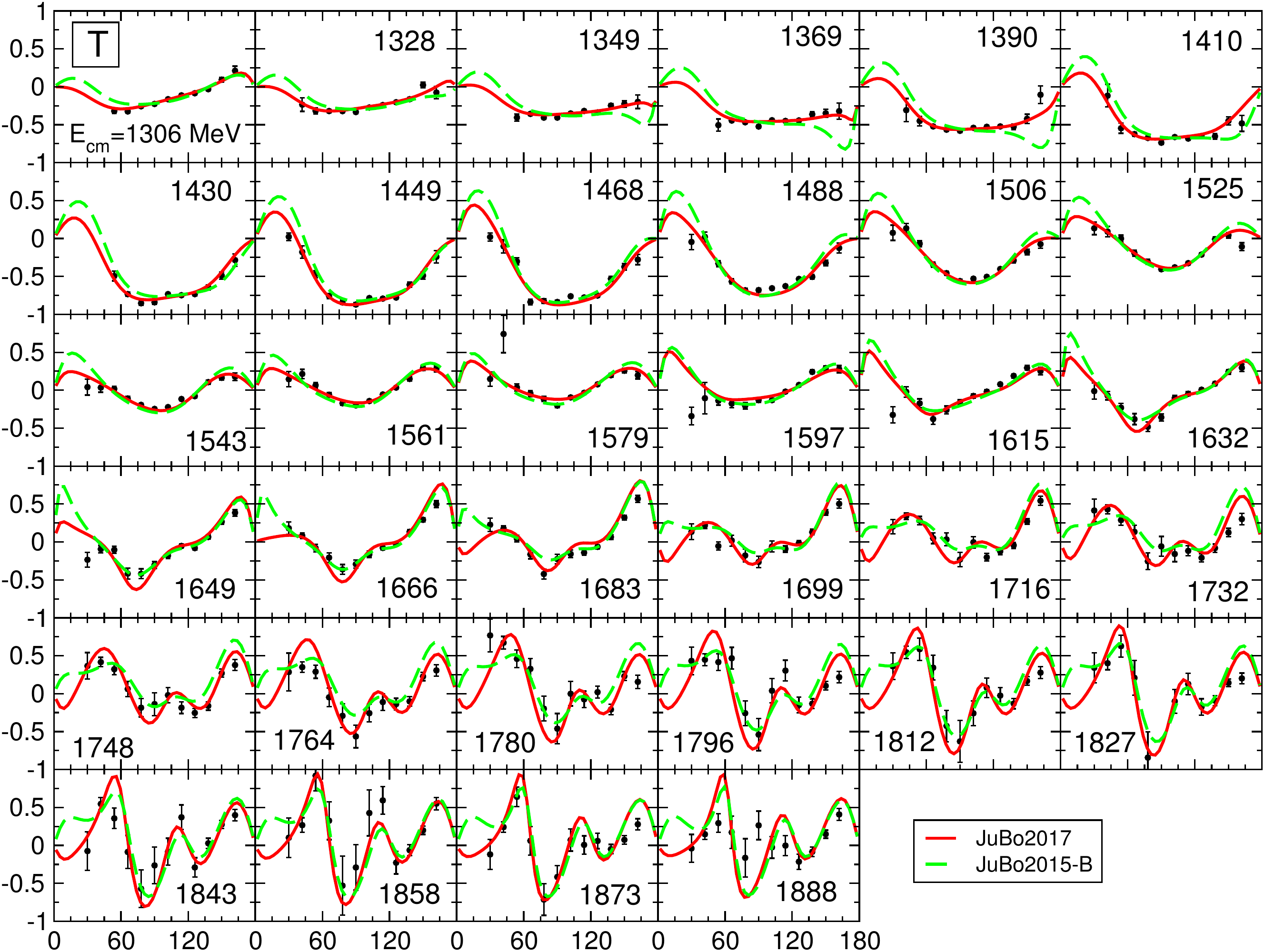} \\
\includegraphics[width=1\linewidth]{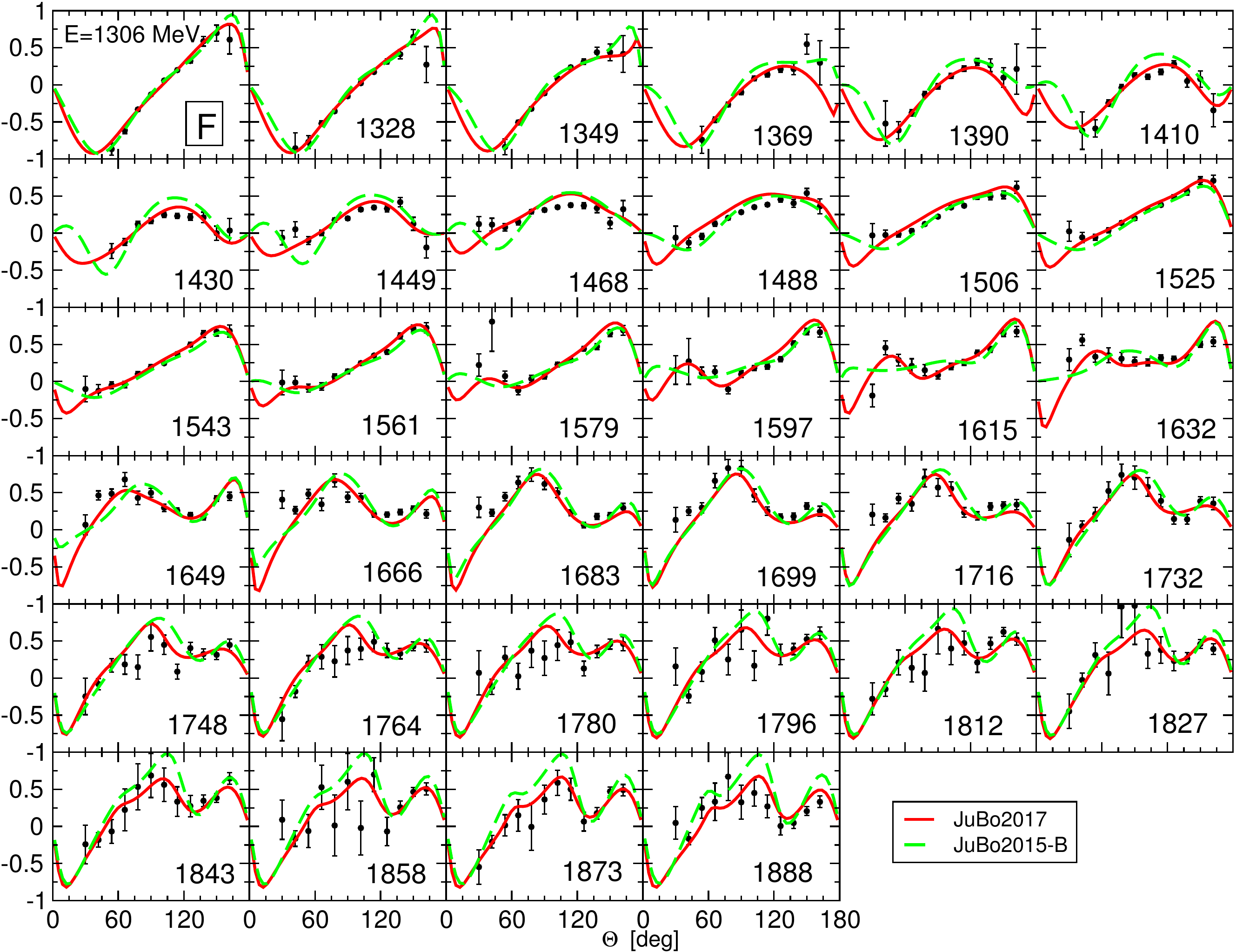} 
\end{center}
\caption{Polarization observables $T$ (above) and $F$ (below) of the reaction $\gamma p\to \pi^0p$. Data: A2 and MAMI (Annand {\it et al.}~\cite{Annand:2016ppc}). Solid (red) lines: this solution J\"uBo2017; green (dashed) lines: prediction from J\"uBo2015-B~\cite{Ronchen:2015vfa}.  }
\label{fig:TFpi0p}
\end{figure}

In Fig.~\ref{fig:TFpi0p} we show the recent measurement of the asymmetries $T$ and $F$ in $\gamma p\to \pi^0p$ together with the solution J\"uBo2015-B~\cite{Ronchen:2015vfa}, where the data were not included in the fit, and the fit result of the present solution J\"uBo2017.

\section{Weights applied in the fit to the $\gamma p\to K^+\Lambda$ data}
\label{app:weights}
In Tab.~\ref{tab:weights} we list the different weights applied in the fit to $\gamma p\to K^+\Lambda$ reaction. Weights for other reactions can be supplied on request. The data included in the present study and the strategy for weighting different data sets can be found in Sec.~\ref{sec:data}. In order to achieve a good fit result for all observables over the whole energy range, it can be necessary to apply different weights for certain energies of a given observable instead of one uniform weight for one observable. Improving the data description for a range of energies by using this procedure reduces the need to introduce new $s$-channel poles in the formalism.

\begin{table*}[]
\caption{Weights applied in the fit to the $\gamma p\to K^+\Lambda$ data.}
\begin{center}
\renewcommand{\arraystretch}{1.4}
\begin {tabular}{lcc|ccc} 
\hline\hline
Observable & Energy range [MeV] &\hspace{0.25cm} Weight \hspace{0.25cm} & \hspace{0.2cm} Observable \hspace{0.2cm} & Energy range [MeV] &\hspace{0.25cm} Weight \hspace{0.25cm}\\ \hline
$d\sigma/d\Omega$ & $E_{\rm cm}<1800$ & 4 & $C_{z^\prime}$ & $E_{\rm cm}<$2080 & 10\\
				&1800$< E_{\rm cm}<1910$ & 28 && $E_{\rm cm}>$2080 & 100\\
				&$E_{\rm cm}>$1910 & 4 & &&\\ \hline
$P$ & $E_{\rm cm}<1800$ & 50 & $O_{x^\prime}$ & $E_{\rm cm}<$1726 & 10  \\
	& $E_{\rm cm}>1800$ & 20 & & 1726$<E_{\rm cm}<$1800 & 40 \\ 
	& && &  $E_{\rm cm}>$1800 & 30 \\ \hline
$\Sigma$~\cite{ Hicks:2007zz, Lleres:2007tx, Sumihama:2005er, Zegers:2003ux}& $E_{\rm cm}<2300$ & 4 & $O_{z^\prime}$& $E_{\rm cm}<$1700 & 30\\
$\Sigma$~\cite{Paterson:2016vmc}& $E_{\rm cm}<2000$ & 17 & &  1700$<E_{\rm cm}<$1800 & 10\\
 					& 2000$<E_{\rm cm}<2140$ & 68 & &  $E_{\rm cm}>$1800 & 50\\
 					& $E_{\rm cm}>2140$ & 510&&\\ \hline
$T$~\cite{Lleres:2008em, Althoff:1978qw}& $E_{\rm cm}<2300$ &1 & $O_x$ & $E_{\rm cm}<$1859 & 20  \\
$T$~\cite{Paterson:2016vmc}& $E_{\rm cm}<2139$ & 11 & &  1859$<E_{\rm cm}<$1919 & 200 \\
						& $E_{\rm cm}>2139$ & 55 & &  1919$<E_{\rm cm}<$2000 & 20\\
				&&& &  $E_{\rm cm}>$2000 & 200 \\		 \hline
$C_{x^\prime}$ & $E_{\rm cm}<1800$ & 12 & $O_z$ & $E_{\rm cm}<$1819 & 18 \\
			&  1800$<E_{\rm cm}<$2035 & 96 & &  1819$<E_{\rm cm}<$1940 & 270\\
			&  2035$<E_{\rm cm}<$2297 & 144 & &  1940$<E_{\rm cm}<$2139 & 18 \\
			&  $E_{\rm cm}>$2297 & 300 & &  $E_{\rm cm}>$2139 & 270 \\ \hline\hline
\end {tabular}
\end{center}
\label{tab:weights}
\end{table*}

\section{Beam-recoil asymmetries}

In Ref.~\cite{Ronchen:2014cna} the observables used in the present study are defined in terms of four amplitudes $F_i$ that are connected to the photoproduction amplitude $\mathcal M$ of Eq.~(\ref{eq:m2}) via a multipole decomposition. Note that our $F_i$ slightly differ from the CGLN amplitudes of Ref.~\cite{Chew:1957tf}. In the following we give the definition of the beam-recoil polarizations $O_x$ and $O_z$ which were not presented in Ref.~\cite{Ronchen:2014cna}:
\begin{eqnarray}
\frac{d\sigma}{d\Omega}O_x&=&-{\rm Im}\left[(F^{}_2-F^{}_3)F^*_1+(F^{}_2\sin^2\theta +F^{}_1\cos\theta )F^*_4\right]
\nonumber \\
&\times& \sin\theta \ , \nonumber \\ 
\frac{d\sigma}{d\Omega}O_z&=&-{\rm Im}\left[F^*_1F^{}_4-(F^{}_3+F^{}_4\cos\theta)F^*_2\right]\sin^2\theta \,. \nonumber
\end{eqnarray}
For all other observables, the decomposition of the $F_i$ into multipoles and definitions of the coordinate frame, the reader is referred to Ref.~\cite{Ronchen:2014cna}.

\newpage

\end{document}